\documentclass[useAMS,usegraphicx,usenatbib]{mn2e}

\newcommand{\Msun}{\ensuremath{\,{\rm M}_\odot}}           
\newcommand{\Rsun}{\ensuremath{\,{\rm R}_\odot}}           
\newcommand{\Teff}{\ensuremath{T_{\rm eff}}}               
\newcommand{\Mjup}{\ensuremath{\,{\rm M}_{\rm Jup}}}       
\newcommand{\Rjup}{\ensuremath{\,{\rm R}_{\rm Jup}}}       
\newcommand{\ms}{\,m\,s$^{-1}$}                            
\newcommand{\mss}{\,m\,s$^{-2}$}                           
\newcommand{\mc}[1]{\multicolumn{2}{c}{#1}}
\newcommand{\as}{\ensuremath{^{\prime\prime}}}              
\newcommand{\am}{\ensuremath{^\prime}}                      
\newcommand{\er}[3]{\ensuremath{#1^{+#2}_{-#3}}}

\newcommand{\erm}[3]{\mc{\ensuremath{#1^{+#2}_{-#3}}}}
\newcommand{\ermm}[5]{\mc{\ensuremath{ {#1\,^{+#2}_{-#3}}\,^{+#4}_{-#5} }}}
\newcommand{\MoH}{\ensuremath{\left[\frac{\rm M}{\rm H}\right]}}        
\newcommand{\pjup}{\ensuremath{\,\rho_{\rm Jup}}}          
\newcommand{\psun}{\ensuremath{\,\rho_\odot}}              

\title[High-precision defocussed photometry of WASP-4]
      {High-precision photometry by telescope defocussing. II. The transiting planetary system WASP-4\thanks{Based on data collected by MiNDSTEp with the Danish 1.54\,m telescope at the ESO La Silla Observatory}}

\author[Southworth et al.]
       {John Southworth\,$^1$\thanks{E-mail: jkt@astro.keele.ac.uk},
        T.\ C.\ Hinse\,$^{2,3}$,          
        M.\ J.\ Burgdorf\,$^4$,           
        M.\ Dominik\,$^5$\thanks{Royal Society University Research Fellow}, 
        A.\ Hornstrup\,$^6$,              
        \newauthor
        U.\ G.\ J{\o}rgensen\,$^2$,       
        C.\ Liebig\,$^7$,                 
        D.\ Ricci\,$^8$,                  
        C.\ C.\ Th\"one$^{9,10}$,         
        T.\ Anguita\,$^7$,                
        V.\ Bozza\,$^{11,12}$,            
        \newauthor
        S.\ Calchi Novati\,$^{11,12}$,    
        K.\ Harps{\o}e\,$^2$,             
        L.\ Mancini\,$^{11,12}$,          
        G.\ Masi\,$^{13}$,                
        M.\ Mathiasen\,$^2$,              
        \newauthor
        S.\ Rahvar\,$^{14}$,              
        G.\ Scarpetta\,$^{11,12}$,        
        C.\ Snodgrass\,$^{15}$,           
        J.\ Surdej\,$^8$,                 
        M.\ Zub$^7$                       
        \\
        $^1$\,Department of Physics, University of Warwick, Coventry, CV4 7AL, UK \\
        $^2$\,Niels Bohr Institute, University of Copenhagen, Juliane Maries Vej 30, Copenhagen \O, 2100, Denmark \\
        $^3$\,Armagh Observatory, College Hill, Armagh, BT61 9DG, Northern Ireland, UK \\
        $^4$\,Deutsches SOFIA Institut, Universitaet Stuttgart, Pfaffenwaldring 31, 70569 Stuttgart, Germany \\
        $^5$\,SUPA, University of St Andrews, School of Physics \& Astronomy, North Haugh, St Andrews, KY16 9SS, UK \\
        $^6$\,National Space Institute, Technical University of Denmark,
                                          Juliane Maries Vej 30, Copenhagen \O, 2100, Denmark  \\
        $^7$\,Astronomisches Rechen-Institut, Zentrum f\"ur Astronomie, Universit\"at Heidelberg,
                                          M\"onchhofstrasse 12-14, 69120 Heidelberg, Germany \\
        $^8$\,Institut d'Astrophysique et de G\'eophysique, Universit\'e de Li\`ege, 4000 Li\`ege, Belgium \\
        $^9$\,Dark Cosmology Centre, Niels Bohr Institute, University of Copenhagen,
                                          Juliane Maries Vej 30, Copenhagen \O, 2100, Denmark \\
        $^{10}$\,INAF, Osservatorio Astronomico di Brera, 23807 Merate, Italy \\
        $^{11}$\,Dipartimento di Fisica ``E. R. Caianiello'', Universit\`a di Salerno, Baronissi, Italy \\
        $^{12}$\,Instituto Nazionale di Fisica Nucleare, Sezione di Napoli, Italy \\
        $^{13}$\,Bellatrix Observatory, Centre for Backyard Astrophysics, Ceccano (FR), Italy \\
        $^{14}$\,Department of Physics, Sharif University of Technology, Tehran, Iran \\
        $^{15}$\,European Southern Observatory, Casilla 19001, Santiago 19, Chile
        }

\begin{document} \maketitle 

\begin{abstract}
We present and analyse light curves of four transits of the Southern hemisphere extrasolar planetary system WASP-4, obtained with a telescope defocussed so the radius of each point spread function was 17\as\ (44 pixels). This approach minimises both random and systematic errors, allowing us to achieve scatters of between 0.60 and 0.88 mmag per observation over complete transit events. The light curves are augmented by published observations and analysed using the {\sc jktebop} code. The results of this process are combined with theoretical stellar model predictions to derive the physical properties of the WASP-4 system. We find that the mass and radius of the planet are $M_{\rm b} = 1.289 \,^{+0.090}_{-0.090} \,^{+0.039}_{-0.000}$ \Mjup\ and $R_{\rm b} = 1.371 \,^{+0.032}_{-0.035} \,^{+0.021}_{-0.000}$ \Rjup, respectively (statistical and systematic uncertainties). These quantities give a surface gravity and density of $g_{\rm b} = 17.03 \,^{+0.97}_{-0.54}$ \mss\ and $\rho_{\rm b} = 0.500\,^{+0.032}_{-0.021} \,^{+0.000}_{-0.008}$ \pjup, and fit the trends for short-period extrasolar planets to have relatively high masses and surface gravities. WASP-4 is now one of the best-quantified transiting extrasolar planetary systems, and significant further progress requires improvements to our understanding of the physical properties of low-mass stars.
\end{abstract}



\section{Introduction}

The discovery of the first extrasolar planet \citep{MayorQueloz95nat} heralded the arrival of a new field of astronomical research: the investigation of how planets form and evolve. To date over 350 extrasolar planets have been discovered, most through the radial velocity motion of their parent stars. The resulting sample is statistically valuable, but only limited information can be extracted for individual objects. The existence of transiting extrasolar planets (TEPs) provides the solution to this problem, as for these objects it is possible to determine their masses, radii and temperatures, and hence surface gravities, densities and to some extent chemical compositions. At present about fifty TEPs are known, but the discovery rate is increasing fast.

\begin{table*} \centering
\caption{\label{tab:obslog} Log of the observations presented in this work. $N_{\rm obs}$ is the number of observations and `Moon' is the fractional illumination of the Moon at the midpoint of the transit.}
\begin{tabular}{lcccccccc} \hline \hline
Date & Start time (UT) & End time (UT) & $N_{\rm obs}$ & Exposure time (s) & Filter & Airmass & Moon & Scatter (mmag)\\
\hline
2008 08 19 & 04:47 & 10:24 & 134 & 120.0 & $R_C$ & 1.11 $\to$ 1.03 $\to$ 1.54 & 0.936 & 0.84 \\  
2008 08 23 & 04:30 & 10:19 & 126 & 120.0 & $R_C$ & 1.10 $\to$ 1.03 $\to$ 1.59 & 0.580 & 0.88 \\  
2008 09 23 & 01:15 & 05:04 &  88 & 120.0 & $R_C$ & 1.27 $\to$ 1.03 $\to$ 1.04 & 0.402 & 0.75 \\  
2008 10 01 & 00:57 & 05:27 & 103 & 120.0 & $R_C$ & 1.23 $\to$ 1.03 $\to$ 1.10 & 0.035 & 0.60 \\  
\hline \hline \end{tabular} \end{table*}

Our ability to determine the physical properties of a TEP is very dependent on the quality of the available observational data. \citet{Me08mn,Me09mn} presented an exhaustive study of fourteen TEPs, to provide an homogeneously measured set of physical properties which can form the basis of statistical studies of these objects. This work highlighted the fact that measurements of the properties of TEPs depend critically on the quality of their transit light curves. The limiting factor for such observations is that the transits are very shallow (only a few percent of the star's light is lost) so it is only possible to measure transit shapes well if observational errors are very low.

In Paper\,I \citep{Me+09} we investigated telescope defocussing as a way to minimise the random and particularly the systematic errors which afflict astronomical time-series photometry. This technique has a long history but has only very recently been regularly applied to charge-coupled device (CCD) observations (see references in Paper\,I). The most important advantage is that flat-fielding errors decrease according to the square-root of the number of pixels in a point spread function (PSF), so the effects of flat-fielding can be lowered by orders of magnitude compared to focussed observations. Changes in atmospheric seeing and telescope pointing affect photometry through flat-fielding errors, so are similarly diminished. Another benefit is that longer exposure times are possible without saturating the CCD, which can thus be read out less often. The main downside is that more background light is collected, but for many situations this is unimportant. The signal-to-noise calculations presented in Paper\,I show that the optimum defocus, in the sense of achieving the best signal to noise per unit time, is surprisingly large and can lead to exposure times of many hundreds of seconds even during periods of high sky brightness. The limiting factor is actually the need to adequately sample the temporal variation of transit events.

In this work we present defocussed photometric observations of four transits of the Southern TEP WASP-4, which was discovered by the SuperWASP collaboration \citep{Wilson+08apj}. This object has since been studied by \citet{Gillon+09aa}, who used the Very Large Telescope (VLT) to observe one transit, and \citet{Winn+09aj}, who measured two transits with the Magellan Baade telescope. These two studies both used defocussing to improve their photometric precision, and the larger telescope apertures (8.2\, and 6.5\,m) also allowed a relatively high observing cadence. Our observations used a 1.5\,m telescope, have a photometric precision comparable to those of \citet{Gillon+09aa} and \citet{Winn+09aj}, and cover more transit events but at a lower sampling rate. In this work we will analyse these three sets of transit observations, plus those obtained by \citet{Wilson+08apj} using the 1.2\,m Euler telescope. Our methods are identical to those used by \citet{Me08mn,Me09mn}, so are fully homogeneous with the physical properties of fourteen TEPs determined in these works.


\section{Observations and data reduction}

\begin{figure} \includegraphics[width=0.48\textwidth,angle=0]{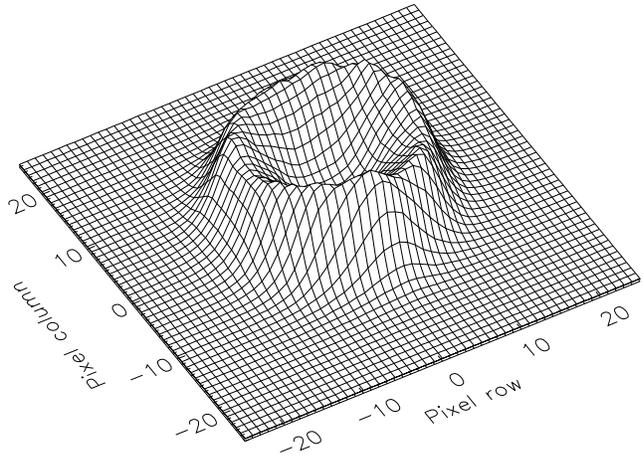}
\caption{\label{fig:psf} Surface plot of the PSF of WASP-4 in an image taken at
random from the observing sequence on the night of 2008 October 1st. The $x$ and
$y$ axes are in pixels. The lowest and highest counts are 587 and 25823 electrons,
respectively, and the $z$ axis is on a linear scale.} \end{figure}

We observed four transits of WASP-4 using the 1.54\,m Danish%
\footnote{Information on the 1.54\,m Danish Telescope and DFOSC can be found at
{\tt http://www.eso.org/sci/facilities/lasilla/telescopes/d1p5/}}
Telescope at ESO La Silla and the DFOSC focal-reducing imager. This setup yielded a full field of view of 13.7\am$\times$13.7\am\ and a plate scale of 0.39\as\,pixel$^{-1}$. The CCD was windowed down to 1200$\times$1000 pixels to decrease the readout time from approximately 90\,s to 30\,s. An observing log is given in Table\,\ref{tab:obslog}.  All observations were done through the Cousins $R$ filter, using exposure times of 120\,s. The amount of defocussing was adjusted until the peak counts per pixel from WASP-4 were roughly 25\,000 above the sky background, resulting in a doughnut-shaped PSF with a diameter of about 44 pixels (17\as). The pointing of the telescope was maintained using autoguiding, and we did not change the amount of defocussing during an observing sequence. An example PSF is shown in Fig.\,\ref{fig:psf}.

Several images were also taken with the telescope properly focussed, and were used to verify that there were no faint stars within the defocussed PSF of WASP-4 which might dilute the transit depth. We found that the closest detectable star is much fainter and at a distance of 31\as\ (78 pixels), so the edge of its PSF is separated from that of WASP-4 by 34 pixels in our defocussed images. The closest star of similar to or greater brightness than WASP-4 lies 80\as\ away. We conclude that no stars interfere with the PSF of WASP-4.

Data reduction was performed in the same way as in Paper\,I, so we only give a summary here. We used a custom pipeline written in the {\sc idl}%
\footnote{The acronym {\sc idl} stands for Interactive Data Language and is a trademark of ITT Visual
Information Solutions. For further details see {\tt http://www.ittvis.com/ProductServices/IDL.aspx}.}
programming language and using the {\sc daophot} package \citep{Stetson87pasp} to perform aperture photometry with the {\sc aper} routine (distributed as part of the {\sc astrolib}%
\footnote{The {\sc astrolib} subroutine library is distributed by NASA. For further details see {\tt http://idlastro.gsfc.nasa.gov/}.}
library). Apertures were placed by eye then fixed throughout each observing sequence. The aperture which gave the most precise photometry was of radius 26 pixels, with a sky annulus of 33--50 pixels; the shape of the light curve is not sensitive to the aperture sizes.

\begin{figure} \includegraphics[width=0.48\textwidth,angle=0]{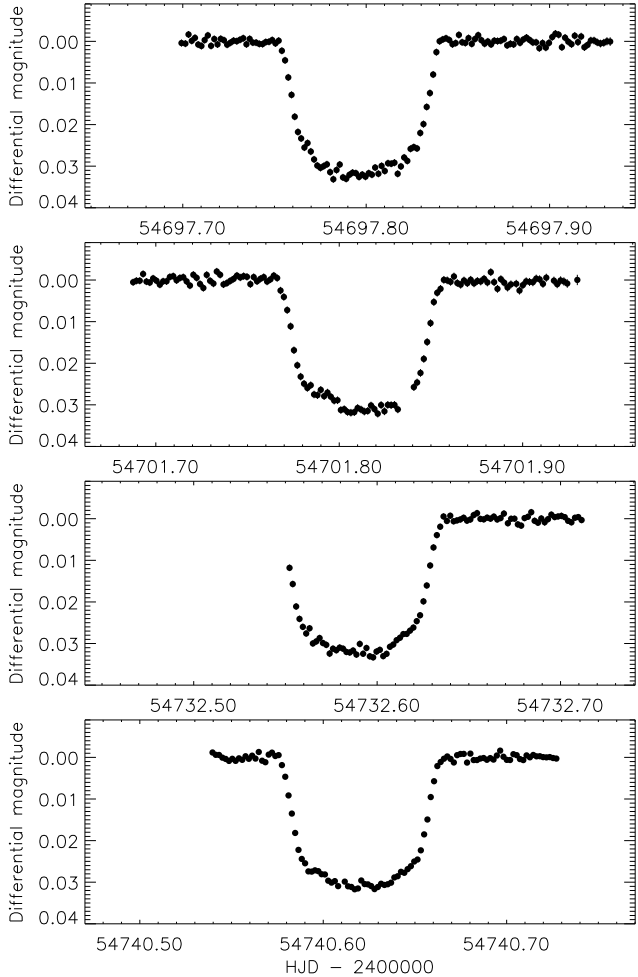}
\caption{\label{fig:plotlc} Final light curves of WASP-4 from our four nights
of observations. The error bars have been scaled to give $\chi^2_{\ \nu} = 1.0$
for each night, and are mostly smaller than the symbol sizes.} \end{figure}

\begin{table} \centering
\caption{\label{tab:lc} Excerpts of the light curve of WASP-4. The full dataset
will be made available in the electronic version of this paper and at the CDS.}
\begin{tabular}{lcr} \hline \hline
HJD & Differential magnitude & Uncertainty \\
\hline
2454697.700127 &  $0.00040$ & $0.00086$ \\
2454697.702257 &  $0.00055$ & $0.00085$ \\[2pt]
2454701.688470 &  $0.00052$ & $0.00085$ \\
2454701.690241 &  $0.00018$ & $0.00085$ \\[2pt]
2454732.553043 &  $0.01181$ & $0.00079$ \\
2454732.554814 &  $0.01570$ & $0.00079$ \\[2pt]
2454740.540454 & $-0.00117$ & $0.00049$ \\
2454740.542225 & $-0.00061$ & $0.00049$ \\
\hline \hline \end{tabular} \end{table}

\begin{table} \begin{center}
\caption{\label{tab:minima} Literature times of minimum light of
WASP-4 and their residuals versus the ephemeris derived in this work.
\newline {\bf References:} (1) \citet{Gillon+09aa};
(2) \citet{Winn+09aj}; (3) This work.}
\begin{tabular}{l r r r} \hline \hline
Time of minimum (HJD)  &    Cycle    &  $O-C$ value  &  Reference  \\
\                      &    number   &  (HJD)        &             \\
\hline
\er{2453963.1086}{0.0025}{0.0021}    & $-$549.0  &    0.00021 & 1  \\
\er{2454364.5757}{0.0021}{0.0033}    &  $-$249.0 & $-$0.00214 & 1  \\
\er{2454368.59266}{0.00025}{0.00027} &  $-$246.0 &    0.00012 & 1  \\
\er{2454371.26738}{0.00097}{0.00087} &  $-$244.0 & $-$0.00162 & 1  \\
\er{2454396.69548}{0.00015}{0.00026} &  $-$225.0 &    0.00008 & 1  \\
2454697.797489 $\pm$ 0.000055        &     0.0   &    0.00000 & 2  \\
2454748.650490 $\pm$ 0.000072        &    38.0   &    0.00021 & 2  \\
2454697.79748 $\pm$ 0.00010          &     0.0   & $-$0.00001 & 3  \\
2454701.81218 $\pm$ 0.00013          &     3.0   & $-$0.00000 & 3  \\
2454732.59118 $\pm$ 0.00013          &    26.0   & $-$0.00033 & 3  \\
2454740.620821 $\pm$ 0.000061        &    32.0   & $-$0.00008 & 3  \\
\hline \hline \end{tabular} \end{center} \end{table}

Five comparison stars were measured on each image, and subsequently checked for short-period variability. Many differential-magnitude light curves were calculated for WASP-4 and between comparison stars. Slow variations in brightness were observed for almost all of these, and are attributable to atmospheric effects. We therefore defined time intervals outside transit events and fitted straight lines to the magnitudes to rectify the light curve shape. Simultaneously with this procedure, the comparison stars were combined into a weighted ensemble which produced the final light curve with the least scatter in the intervals outside transit. We found that including flat-field corrections provided only a slight improvement in every case, as the telescope pointing was good, and that accounting for the CCD bias had a negligible effect. We have applied bias and flat-field corrections to generate our final light curves, which are shown in Fig.\,\ref{fig:plotlc} and reproduced in Table\,\ref{tab:lc}. The scatter in the final light curves varies from 0.60 to 0.88 mmag, and the most precise data were obtained in dark time when the sky background was low.


\section{Light curve analysis}      \label{sec:lc}

\subsection{Period determination}

\begin{figure*} \includegraphics[width=\textwidth,angle=0]{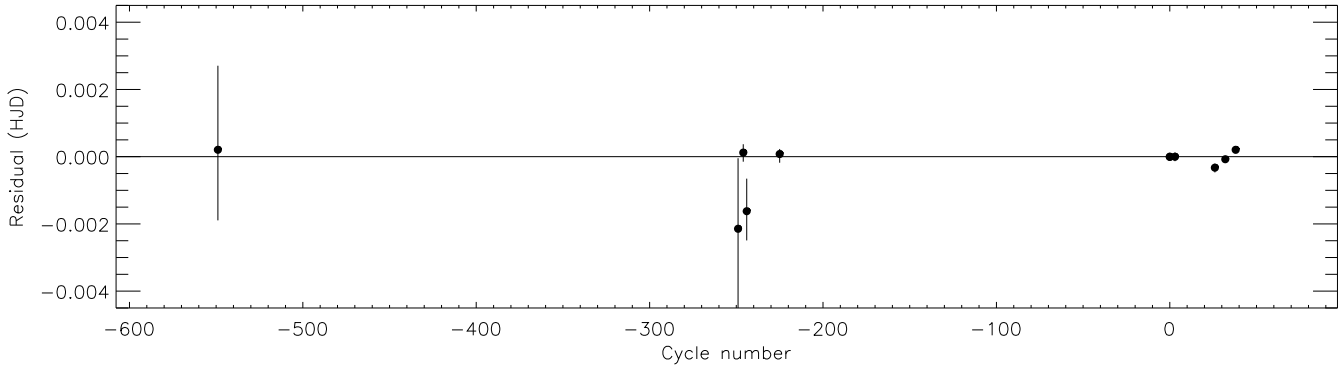}
\caption{\label{fig:minima} Plot of the residuals of the times of
mid-transit of WASP-4 versus a linear ephemeris. Some error bars
are smaller than the symbol sizes.} \end{figure*}

The first step in defining the physical properties of the WASP-4 system is an accurate orbital ephemeris. We have obtained individual times of mid-transit by analysing the data from our four nights separately, using the {\sc jktebop} code and Monte Carlo simulations (see below). These have been supplemented by the two transit timings given by \citet{Winn+09aj} and the five listed by \citet{Gillon+09aa} found from their `prayer-bead' analysis.

We have assigned cycle numbers to the times of minimum light, taking as reference epoch the midpoint of our first transit (which by coincidence is the same transit event as the first one observed by \citealt{Winn+09aj}), and fitted a straight line to the transit timings. We find an orbital ephemeris of
$$ T_0 = {\rm HJD} \,\, 2\,454\,697.797488 (31) \, + \, 1.33823150 (61) \times E $$
where $E$ is the number of orbital cycles after the reference epoch and quantities in parentheses denote the uncertainty in the final digit of the preceding number. The reduced $\chi^2$ of the fit is $\chi_{\nu}^{\,2} = 1.35$, which suggests the possibility that the orbital period is not constant. The transit timings are collected in Table\,\ref{tab:minima} along with the residuals of the straight-line fit, which plotted in Fig.\,\ref{fig:minima}.

In the course of measuring the times of minimum light given by our data, we found that the absolute sizes of the observational error bars delivered by our pipeline (which originate from the {\sc aper} algorithm) are not reliable. We have rescaled the error bars by multiplying them by $\sqrt{\chi_{\nu}^{\,2}}$ to give $\chi_{\nu}^{\,2} = 1.0$ \citep{Bruntt+06aa,Me++07aa}, treating each night individually.

\subsection{Light curve modelling}

%

\begin{figure} \includegraphics[width=0.48\textwidth,angle=0]{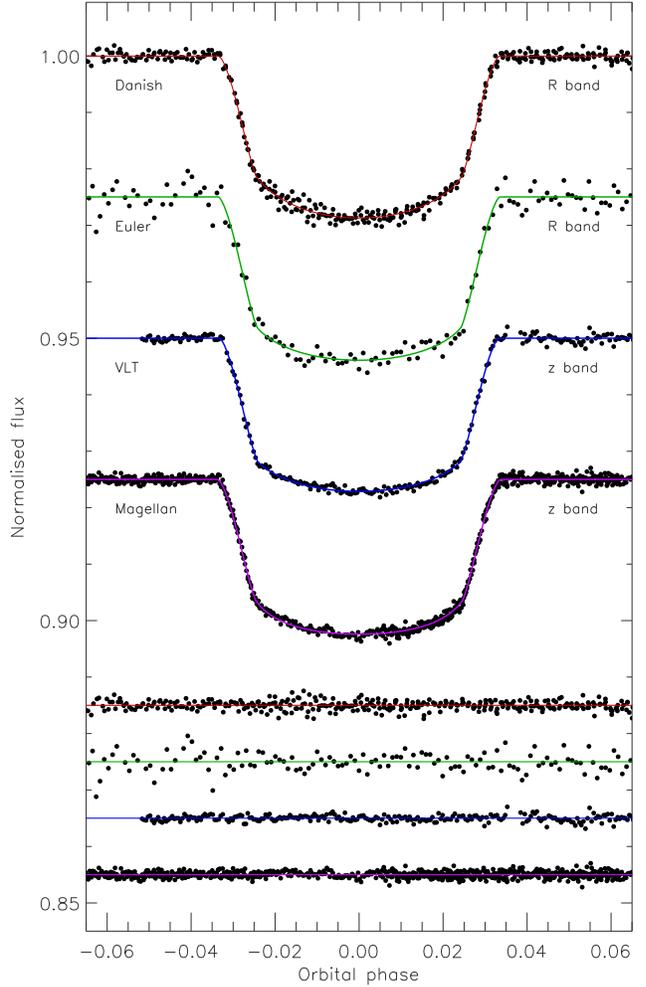}
\caption{\label{fig:lcfit} \label{fig:lcfitpub} Phased light curves of WASP-4,
compared to the best fits found using {\sc jktebop} and the quadratic LD law
with both LD coefficients included as fitted parameters. Top is from the Danish
telescope (this work); second is from the Euler telescope \citep{Wilson+08apj};
third from the VLT \citep{Gillon+09aa}; last is from Magellan \citep{Winn+09aj}.
The residuals of the fits are plotted at the bottom of the figure, offset from
zero.} \end{figure}

We have analysed the light curves using an identical approach to that discussed in detail by \citet{Me08mn}. We therefore retain homogeneity with the fourteen TEPs analysed in that work, as well as with our study of WASP-5 (Paper\,I). The light curves were modelled using the {\sc jktebop}%
\footnote{{\sc jktebop} is written in {\sc fortran77} and the source code is available at {\tt http://www.astro.keele.ac.uk/$\sim$jkt/}}
code \citep{Me++04mn,Me++04mn2}, which is based on the {\sc ebop} program originally developed for eclipsing binary star systems \citep{PopperEtzel81aj,Etzel81conf,NelsonDavis72apj}. This simulates the two components using biaxial spheroids, so allows departures from sphericity. The parameter which governs the shapes of the components is the mass ratio; we have adopted the value 0.0013 and find that even large changes have a negligible effect on the solution.

The main results from the light curve analysis are the best-fitting orbital inclination, $i$, and fractional radii of the star and planet, $r_{\rm A} = \frac{R_{\rm A}}{a}$ and $r_{\rm b} = \frac{R_{\rm b}}{a}$, where $R_{\rm A}$ and $R_{\rm b}$ are the absolute radii of the components and $a$ is the orbital semimajor axis. $r_{\rm A}$ and $r_{\rm b}$ are actually parameterised as their sum and ratio, ($r_{\rm A}$$+$$r_{\rm b}$) and $k = \frac{r_{\rm A}}{r_{\rm b}}$, because these two quantities are only weakly correlated for a wide variety of transiting planet and other eclipsing binary systems.

We included the limb darkening (LD) of the star using five different LD laws (see \citealt{Me08mn} for their definition) with the coefficients of these laws either fixed at theoretically predicted values, included as fitted parameters, or perturbed around theoretical values. The theoretical coefficients were obtained by bilinear interpolation in the tabulations of \citet{Vanhamme93aj}, \citet{Claret00aa,Claret04aa2} and \citet{ClaretHauschildt03aa}.

The uncertainties of the fitted parameters were assessed using 1000 Monte Carlo simulations for each solution \citep{Me+04mn3,Me+05mn}. We have also tested for correlated (`red') noise using the residual-permutation method \citep{Jenkins++02apj}, finding a small but significant amount which is probably of instrumental origin. However, it is possible that it is caused by spots on the surface of the parent star, based on the shapes of our transit light curves at HJDs 2454701.79 and 2454732.61 (Fig.\,\ref{fig:plotlc}). The other two transits, one of which was also observed by \citet{Winn+09aj}, are seemingly unaffected.

At the request of the referee we have calculated the parameter $\beta$, a measure of the systematic noise contribution \citep{Winn+07aj2,Gillon+06aa,Winn+08apj2}, for comparison with other works. We evaluated $\beta$ values for individual transits and for groups of between two and ten datapoints. The average values of $\beta$ over these groups are 1.03, 1.40, 1.41 and 1.14 for our four observing sequences. We emphasise that we do not use these numbers in our analysis, but account for systematics using the residual permutation approach instead.


\begin{table*} \caption{\label{tab:lcfit1} Parameters of the {\sc jktebop} best
fits of the light curve of WASP-4 for different LD laws with the coefficients
fixed at theoretically predicted values. For each part of the table the upper
quantities are fitted parameters and the lower quantities are derived parameters.
$T_0$ is given as HJD $-$ 2454000.0.}
\begin{tabular}{l r@{\,$\pm$\,}l r@{\,$\pm$\,}l r@{\,$\pm$\,}l r@{\,$\pm$\,}l r@{\,$\pm$\,}l}
\hline \hline
\                     & \mc{Linear LD law}                   & \mc{Quadratic LD law}                & \mc{Square-root LD law}              & \mc{Logarithmic LD law}              & \mc{Cubic LD law}                    \\
\hline
$r_{\rm A}+r_{\rm b}$ & \erm{0.2103}{0.0021}{0.0006}         & \erm{0.2097}{0.0018}{0.0006}         & \erm{0.2097}{0.0020}{0.0006}         & \erm{0.2099}{0.0021}{0.0005}         & \erm{0.2092}{0.0020}{0.0005}         \\
$k$                   & \erm{0.15245}{0.00045}{0.00029}      & \erm{0.15345}{0.00033}{0.00023}      & \erm{0.15322}{0.00040}{0.00025}      & \erm{0.15316}{0.00039}{0.00024}      & \erm{0.15377}{0.00037}{0.00026}      \\
$i$ (deg.)            & \erm{89.84}{ 0.61}{ 1.04}            & \erm{89.82}{ 0.59}{ 0.91}            & \erm{89.97}{ 0.52}{ 1.04}            & \erm{89.97}{ 0.40}{ 1.17}            & \erm{89.73}{ 0.56}{ 0.97}            \\
$u_{\rm A}$           & \mc{ 0.60 fixed}                     & \mc{ 0.40 fixed}                     & \mc{ 0.25 fixed}                     & \mc{ 0.70 fixed}                     & \mc{ 0.50 fixed}                     \\
$v_{\rm A}$           & \mc{ }                               & \mc{ 0.25 fixed}                     & \mc{ 0.50 fixed}                     & \mc{ 0.23 fixed}                     & \mc{ 0.10 fixed}                     \\
$T_0$                 & \erm{365.916894}{0.000047}{0.000045} & \erm{365.916893}{0.000041}{0.000047} & \erm{365.916894}{0.000041}{0.000047} & \erm{365.916894}{0.000045}{0.000049} & \erm{365.916894}{0.000043}{0.000046} \\
\hline
$r_{\rm A}$           & \erm{0.1825}{0.0018}{0.0005}         & \erm{0.1818}{0.0015}{0.0005}         & \erm{0.1819}{0.0017}{0.0005}         & \erm{0.1820}{0.0018}{0.0004}         & \erm{0.1813}{0.0016}{0.0004}         \\
$r_{\rm b}$           & \erm{0.02782}{0.00035}{0.00011}      & \erm{0.02790}{0.00029}{0.00010}      & \erm{0.02786}{0.00034}{0.00009}      & \erm{0.02787}{0.00035}{0.00008}      & \erm{0.02788}{0.00032}{0.00009}      \\
$\sigma$ ($m$mag)     & \mc{ 0.8792}                         & \mc{ 0.8576}                         & \mc{ 0.8558}                         & \mc{ 0.8555}                         & \mc{ 0.8553}                         \\
$\chi^2_{\rm \ red}$  & \mc{ 1.4308}                         & \mc{ 1.3189}                         & \mc{ 1.3187}                         & \mc{ 1.3164}                         & \mc{ 1.3107}                         \\
\hline \hline \end{tabular} \end{table*}

\begin{table*} \caption{\label{tab:lcfit2} Parameters of the {\sc jktebop} best
fits of the light curve of WASP-4 for different LD laws with all LD coefficients
included as fitted parameters. For each part of the table the upper quantities
are fitted parameters and the lower quantities are derived parameters.
$T_0$ is given as HJD $-$ 2454000.0.}
\begin{tabular}{l r@{\,$\pm$\,}l r@{\,$\pm$\,}l r@{\,$\pm$\,}l r@{\,$\pm$\,}l r@{\,$\pm$\,}l}
\hline \hline
\                     & \mc{Linear LD law}                   & \mc{Quadratic LD law}                & \mc{Square-root LD law}              & \mc{Logarithmic LD law}              & \mc{Cubic LD law}                    \\
\hline
$r_{\rm A}+r_{\rm b}$ & \erm{0.2085}{0.0019}{0.0008}         & \erm{0.2088}{0.0023}{0.0011}         & \erm{0.2093}{0.0019}{0.0014}         & \erm{0.2091}{0.0022}{0.0012}         & \erm{0.2093}{0.0022}{0.0015}         \\
$k$                   & \erm{0.15471}{0.00054}{0.00041}      & \erm{0.15428}{0.00103}{0.00075}      & \erm{0.15409}{0.00116}{0.00103}      & \erm{0.15423}{0.00123}{0.00086}      & \erm{0.15426}{0.00130}{0.00106}      \\
$i$ (deg.)            & \erm{89.30}{ 0.76}{ 0.86}            & \erm{90.00}{ 0.60}{ 1.34}            & \erm{89.58}{ 0.67}{ 1.23}            & \erm{89.57}{ 0.75}{ 1.13}            & \erm{89.36}{ 0.92}{ 1.09}            \\
$u_{\rm A}$           & \erm{0.501}{0.013}{0.013}            & \erm{0.476}{0.046}{0.047}            & \erm{0.276}{0.291}{0.294}            & \erm{0.595}{0.117}{0.134}            & \erm{0.493}{0.028}{0.028}            \\
$v_{\rm A}$           & \mc{ }                               & \erm{0.081}{0.108}{0.113}            & \erm{0.412}{0.531}{0.516}            & \erm{0.114}{0.151}{0.169}            & \erm{0.093}{0.144}{0.155}            \\
$T_0$                 & \erm{365.916894}{0.000039}{0.000047} & \erm{365.916885}{0.000052}{0.000047} & \erm{365.916885}{0.000051}{0.000050} & \erm{365.916885}{0.000052}{0.000052} & \erm{365.916884}{0.000050}{0.000053} \\
\hline
$r_{\rm A}$           & \erm{0.1806}{0.0015}{0.0007}         & \erm{0.1809}{0.0019}{0.0010}         & \erm{0.1813}{0.0017}{0.0013}         & \erm{0.1812}{0.0018}{0.0011}         & \erm{0.1814}{0.0018}{0.0013}         \\
$r_{\rm b}$           & \erm{0.02794}{0.00033}{0.00013}      & \erm{0.02791}{0.00053}{0.00011}      & \erm{0.02794}{0.00047}{0.00011}      & \erm{0.02794}{0.00047}{0.00014}      & \erm{0.02797}{0.00053}{0.00017}      \\
$\sigma$ ($m$mag)     & \mc{ 0.8568}                         & \mc{ 0.8518}                         & \mc{ 0.8517}                         & \mc{ 0.8517}                         & \mc{ 0.8518}                         \\
$\chi^2_{\rm \ red}$  & \mc{ 1.3115}                         & \mc{ 1.1364}                         & \mc{ 1.1362}                         & \mc{ 1.1364}                         & \mc{ 1.1365}                         \\
\hline \hline \end{tabular} \end{table*}

The final light curve fits consist of sets of best-fitting parameter values with Monte Carlo uncertainties, for each LD law and for different combinations of fixed/perturbed/fitted LD coefficients. In each case the linear coefficient is represented by $u_{\rm A}$ and the nonlinear coefficient by $v_{\rm A}$. The solutions for all LD coefficients fixed are given in Table\,\ref{tab:lcfit1}, and for all LD coefficients fitted in Table\,\ref{tab:lcfit2}. The latter solutions are clearly better, indicating that the theoretical LD coefficients are not a perfect match to the data. We therefore adopt these solutions, with uncertainties which have been increased to account for the correlated noise discussed in the previous paragraph. A fit to the data is shown in Fig.\,\ref{fig:lcfit}, for which the rms of the residuals is 0.85\,mmag.

In order to maximise the accuracy of the final results, we have also analysed the data obtained by \citet{Gillon+09aa}, \citet{Winn+09aj}, and the Euler Telescope data presented by \citet{Wilson+08apj}. The full sets of solutions of all four light curves are given in the Appendix\footnote{The Appendix is available only in the electronic version of this work.} (Tables \ref{tab:lceuler}, \ref{tab:lcvlt}, \ref{tab:lcmag} and \ref{tab:lcdanish}). In each case we adopted the results for both LD coefficients fitted, and performed Monte Carlo and residual permutation analyses to obtain the final errors. The best fits are shown in Fig.\,\ref{fig:lcfitpub} and the final parameters for each light curve are collected in Table\,\ref{tab:lcfinal}.

Also in Table\,\ref{tab:lcfinal} are the final photometric parameters, obtained by averaging our four sets of results for the individual datasets. We found that the $\chi_\nu^{\,2}$ values for these averages are generally noticeably greater than unity (between 1.1 and 2.3), which suggests that there is additional uncertainty not picked up by our random or systematic error analyses. The error bars in our final photometric parameters have been increased to account for this additional uncertainty. The situation for WASP-4 echoes that for HD\,189733 and HD\,209458 \citep{Me08mn} and should probably be attributed to the effects of starspots. One corollary of this phenomenon is that a detailed study of a TEP cannot rest on high-precision photometry of only a single transit: the additional uncertainty may then exist but not be detectable.

\begin{table*} \centering \caption{\label{tab:lcfinal} Final parameters of the fit
to the light curves of WASP-4. The results of published studies are included for
comparison, and are without error bars if they were not directly quoted quantities.}
 \setlength{\tabcolsep}{4.5pt}
\begin{tabular}{l cccc c ccc}
\hline \hline
Source & This work & This work & This work & This work & This work  & Wilson et al. & Gillon et al. & Winn et al. \\
       & (final)   & (Danish)  & (Euler)   & (VLT)     & (Magellan) & \citeyearpar{Wilson+08apj} & \citeyearpar{Gillon+09aa}& \citeyearpar{Winn+09aj} \\
\hline
$r_{\rm A}+r_{\rm b}$ & \er{0.2106}{0.0010}{0.0012}    & \er{0.2091}{0.0016}{0.0025}    & \er{0.2101}{0.0053}{0.0089}    & \er{0.2100}{0.0023}{0.0022}    & \er{0.2118}{0.0015}{0.0017}    & 0.2190                      & 0.2078                         & 0.2092                         \\
$k$                   & \er{0.15399}{0.00068}{0.00065} & \er{0.1542}{0.0012}{0.0014}    & \er{0.1586}{0.0041}{0.0039}    & \er{0.1541}{0.0012}{0.0010}    & \er{0.15374}{0.00073}{0.00079} & \er{0.1552}{0.0016}{0.0006} & \er{0.15353}{0.00033}{0.00026} & \er{0.15375}{0.00077}{0.00055} \\
$i$ ($^\circ$)        & 88.0 to 90.0                   & \er{89.6}{0.4}{1.0}            & \er{88.8}{1.2}{2.4}            & \er{88.1}{0.7}{1.4}            & \er{88.6}{0.6}{1.2}            & \er{88.59}{1.36}{1.50}      & \er{89.35}{0.64}{0.49}         & \er{88.56}{0.98}{0.46}         \\
$r_{\rm A}$           & \er{0.1825}{0.0011}{0.0010}    & \er{0.1812}{0.0014}{0.0020}    & \er{0.1814}{0.0048}{0.0070}    & \er{0.1820}{0.0021}{0.0018}    & \er{0.1836}{0.0012}{0.0015}    & 0.1895                      & 0.1801                         & \er{0.1827}{0.0017}{0.0005}    \\
$r_{\rm b}$           & \er{0.02812}{0.00022}{0.00014} & \er{0.02794}{0.00019}{0.00057} & \er{0.02876}{0.00056}{0.00172} & \er{0.02805}{0.00045}{0.00046} & \er{0.02822}{0.00028}{0.00030} & 0.02942                     & 0.02764                        & 0.027878                       \\
\hline \hline \end{tabular} \end{table*}

%
%
%
%
%



\section{The physical properties of WASP-4}

\begin{table*} \caption{\label{tab:absdimall} Physical properties for
WASP-4, derived using either an empirical stellar mass--radius relation
or the predictions of different sets of stellar evolutionary models.
Mass, radius, surface gravity and density are denoted by $M$, $R$,
$\log g$ and $\rho$, respectively, subscripted with an `A' for the star
and a `b' for the planet.}
\begin{tabular}{l l r@{\,$\pm$\,}l r@{\,$\pm$\,}l r@{\,$\pm$\,}l r@{\,$\pm$\,}l r@{\,$\pm$\,}l r@{\,$\pm$\,}l}
\hline \hline
\ & \ & \mc{Mass--radius} & \mc{{\sf Padova} models}  & \mc{{\sf Y$^2$} models} & \mc{{\sf Claret} models} \\
\hline
$a$ & (AU)               & 0.022669 & 0.00058       & \erm{0.02294}{0.00050}{0.00040} & \erm{0.02320}{0.00036}{0.00049} & \erm{0.02329}{0.00040}{0.00058} \\
$M_{\rm A}$ & (\Msun)    & 0.867 & 0.067            & \erm{0.898}{0.060}{0.047}       & \erm{0.929}{0.032}{0.057}       & \erm{0.940}{0.048}{0.069}       \\
$R_{\rm A}$ & (\Rsun)    & 0.889 & 0.024            & \erm{0.900}{0.020}{0.016}       & \erm{0.910}{0.015}{0.020}       & \erm{0.914}{0.016}{0.023}       \\
$\log g_{\rm A}$ & [cgs] & 4.478 & 0.012            & \erm{4.483}{0.011}{0.009}       & \erm{4.488}{0.008}{0.011}       & \erm{4.490}{0.009}{0.012}       \\
$\rho_{\rm A}$ & (\psun) & \erm{1.233}{0.020}{0.022}& \erm{1.233}{0.020}{0.022}       & \erm{1.233}{0.020}{0.022}       & \erm{1.233}{0.020}{0.022}       \\
$M_{\rm b}$ & (\Mjup)    & \erm{1.221}{0.093}{0.071}& \erm{1.250}{0.090}{0.056}       & \erm{1.279}{0.082}{0.063}       & \erm{1.289}{0.085}{0.073}       \\
$R_{\rm b}$ & (\Rjup)    & 1.334 & 0.036            & \erm{1.350}{0.032}{0.025}       & \erm{1.365}{0.023}{0.029}       & \erm{1.371}{0.026}{0.035}       \\
$g_{\rm b}$ & (\mss)     & \erm{17.03}{0.97}{0.54}  & \erm{17.03}{0.97}{0.54}         & \erm{17.03}{0.97}{0.54}         & \erm{17.03}{0.97}{0.54}         \\
$\rho_{\rm b}$ & (\pjup) & \erm{0.514}{0.033}{0.023}& \erm{0.508}{0.031}{0.021}       & \erm{0.502}{0.031}{0.020}       & \erm{0.500}{0.032}{0.020}       \\
Age & (Gyr)              & \mc{ }                   & \erm{7.3}{2.3}{4.2}             & \erm{5.5}{3.2}{2.0}             & \erm{6.2}{5.3}{2.6}             \\
\hline \hline \end{tabular} \end{table*}

\begin{table*} \caption{\label{tab:absdim} Final physical properties
for WASP-4. Both statistical and systematic uncertainties are included.
The corresponding results from \citet{Wilson+08apj}, \citet{Gillon+09aa}
and \citet{Winn+09aj} have been included for comparison. The surface
gravity of the planet has no systematic error as it can be specified
by purely observational quantities \citep{Me+04mn3,Me++07mn}.}
\begin{tabular}{l l r@{\,$\pm$\,}l r@{\,$\pm$\,}l r@{\,$\pm$\,}l r@{\,$\pm$\,}l } \hline \hline
\      &                 & \mc{Final result (this work)}                     & \mc{\citet{Wilson+08apj}}  & \mc{\citet{Gillon+09aa}}       & \mc{\citet{Winn+09aj}}   \\
\hline
$a$         & (AU)       & \ermm{0.02329}{0.00050}{0.00035}{0.00058}{0.00000}& 0.0230 & 0.001             & \erm{0.02255}{0.00095}{0.00065}& 0.02340 & 0.00060        \\
$M_{\rm A}$ & (\Msun)    & \ermm{0.940}{0.060}{0.069}{0.042}{0.000}          & \erm{0.8997}{0.077}{0.072} & \erm{0.85}{0.11}{0.07}         & 0.925 & 0.040            \\
$R_{\rm A}$ & (\Rsun)    & \ermm{0.914}{0.020}{0.023}{0.014}{0.000}          & \erm{0.9370}{0.04}{0.03}   & \erm{0.873}{0.036}{0.027}      & 0.912 & 0.013            \\
$\log g_{\rm A}$ & [cgs] & \ermm{4.490}{0.011}{0.012}{0.007}{0.000}          & \erm{4.45}{0.016}{0.029}   & \erm{4.487}{0.019}{0.15}       & 4.4813 & 0.0080          \\
$\rho_{\rm A}$ & (\psun) & \erm{1.233}{0.020}{0.022}                         & \erm{1.094}{0.038}{0.085}  & \erm{1.284}{0.013}{0.019}      & \erm{1.227}{0.011}{0.033}\\
$M_{\rm b}$ & (\Mjup)    & \ermm{1.289}{0.090}{0.073}{0.039}{0.000}          & \erm{1.215}{0.087}{0.079}  & \erm{1.21}{0.13}{0.08}         & 1.237 & 0.064            \\
$R_{\rm b}$ & (\Rjup)    & \ermm{1.371}{0.032}{0.035}{0.021}{0.000}          & \erm{1.416}{0.068}{0.043}  & \erm{1.304}{0.054}{0.042}      & 1.365 & 0.021            \\
$g_{\rm b}$ & (\mss)     & \erm{17.03}{0.97}{0.54}                           & \erm{13.87}{0.75}{1.04}    & \erm{16.29}{0.97}{0.41}        & 16.41 & 0.75             \\
$\rho_{\rm b}$ & (\pjup) & \ermm{0.500}{0.032}{0.021}{0.000}{0.008}          & \erm{0.420}{0.032}{0.044}  & \erm{0.546}{0.039}{0.025}      &0.487 & 0.034             \\
\hline \hline \end{tabular} \end{table*}

The parameters determined from the light curve analysis are not enough in themselves to calculate the physical properties of the components of the WASP-4 system, so additional constraints must be sought from other types of observations and also from theoretical models of stellar evolution. The observational constraints are the velocity amplitude of the parent star, $K_{\rm A} = \er{247.6}{13.9}{6.8}$\ms, plus its spectroscopically-derived effective temperature and surface metal abundance, $\Teff = 5500 \pm 100$\,K and $\MoH = -0.03 \pm 0.09$\,dex \citep{Gillon+09aa}. For theoretical constraints we use  tabulations from the {\sf Claret} \citep{Claret04aa,Claret05aa,Claret06aa2,Claret07aa2}, {\sf Y$^2$} \citep{Demarque+04apjs} and {\sf Padova} \citep{Girardi+00aas} models.

Our approach follows that of \citet{Me09mn}: the above input quantities are specified and the (unknown) velocity amplitude of the {\em planet} is freely adjusted to find the best match between the properties of the star and the predictions of one of the sets of theoretical models. The process is repeated many times whilst varying each input quantity by its 1$\sigma$ uncertainty, building up a complete picture of the solution and its random-error budget \citep{Me++05aa}. There is a clear theoretical dependence arising from the use of stellar models in this procedure, and a lower limit on the resulting systematic errors are inferred by comparison between the results for the three different sets of theoretical predictions. This is only a lower limit because independent models have some physical ingredients in common, for example opacity tables. As an external check we ignore the stellar models and instead force the star to match an empirical mass--radius relation, constructed by \citet{Me09mn} from well-studied eclipsing binary star systems. There is one output parameter which has no dependence on the mass--radius relation or theoretical predictions: the surface gravity of the planet \citep{Me+04mn3,Me++07mn}.

Table\,\ref{tab:absdimall} shows the results of the above process for WASP-4, taking the light curve parameters from Section\,\ref{sec:lc} and the $K_{\rm A}$, \Teff\ and \MoH\ from \citet{Gillon+09aa}. The variations between results for different sets of stellar models are most important for $a$ and $M_{\rm A}$, but for all parameters the systematics are similar to or smaller than the random errors. The results found using the mass--radius relation differ by a similar amount to the systematic errors, being in general smaller. This probably sets an upper limit on the systematic errors in the properties of WASP-4, but further work is still necessary to illuminate the reasons why theoretical models persistently underestimate the radii of low-mass stars \citep{Clausen98conf,TorresRibas02apj,Ribas+08conf}. The {\sf Y$^2$} models yield the most precise of the age estimates, due to the smooth evolution through the main sequence predicted by these models, but the age of WASP-4 remains poorly constrained.

In Table\,\ref{tab:absdim} we give the final physical properties of the WASP-4 system, including both random and systematic error estimates for all parameters, and compare these to those found by other studies. The agreement is in general very acceptable, even though the error bars are small. WASP-4 therefore joins the relatively small group of best-understood TEPs, helped by the large amount of high-quality photometric data available for it. The error budgets we find for its physical properties indicate that an improved value for the velocity amplitude of its parent star would be useful, plus an improved understanding of the stellar \Teff\ and \MoH\ and the systematic errors inherent in the effective temperature scale of low-mass stars.


\section{Summary and conclusions}

Transiting extrasolar planets (TEPs) are vital as our primary source of the physical properties of gas giant planets. The precision with which we can determine the masses and radii of these objects is critically dependent on the observational data we can obtain for them, in particular the quality of the light curves of transit events. Defocussing your telescope allows a much better control of systematic effects, many of which arise through pixel sensitivity variations, and also lowers the amount of observing time lost to CCD readout. Telescope defocussing is rapidly becoming a standard method for observing TEPs, and in this work we have presented defocussed observations of four transits of the WASP-4 system.

We analysed these transit light curves in order to define the photometric parameters for WASP-4 ($r_{\rm A}$, $r_{\rm b}$, $i$ and period), with careful attention paid to the incorporation of stellar limb darkening and to the assessment of random and systematic errors. Three other sets of high-quality light curves (precision better than 2\,mmag per observation) exist in the literature \citep{Wilson+08apj,Gillon+09aa,Winn+09aj}, which we have modelled in the same way. Combining these analyses with our own maximises the quality of the final results and minimises the possibility of systematic errors caused by starspots or instrumental effects. The photometric parameters were then added to the observed stellar velocity amplitude, effective temperature and metal abundance in order to determine the physical properties of the star and planet in the WASP-4 system. We undertook this step using additional constraints from theoretical stellar evolutionary models, and assessed the resulting systematic error by comparing results obtained using different sets of models.

We find that WASP-4\,b has a mass of $M_{\rm b} = 1.289 \,^{+0.090}_{\,-0.073} \,^{\,+0.039}_{\,-0.000}$ \Mjup, a radius of $R_{\rm b} = 1.371 \,^{+0.032}_{\,-0.035} \,^{\,+0.021}_{\,-0.000}$ \Rjup, and a surface gravity of $g_{\rm b} = 17.03 \,^{+0.97}_{-0.54}$ \mss. Its equilibrium temperature is $T_{\rm eq} = 1662 \pm 46$\,K. WASP-4\,b therefore fits in well with the trends for short-period planets to have relatively large masses \citep{Mazeh++05mn} and surface gravities \citep{Me++07mn}. Theoretical models of coreless irradiated gas giant planets \citep{Bodenheimer++03apj, Fortney++07apj, Baraffe++08aa} give predictions of the radius of WASP-4\,b ranging from roughly 1.09 to 1.17\Rjup, and are therefore rejected at beyond the 5$\sigma$ level. Models with a heavy-element core propose smaller radii and are therefore even more discrepant. However, alternative models including an {\it ad hoc} additional kinetic heating source \citep{Bodenheimer++03apj} can satisfactorily explain the properties of WASP-4\,b.


The high quality of the available data for WASP-4, plus the comparative depth of the transits (3\%), means the mass and radius measurements of its planet are among the most precise available, together with CoRoT-Exo-2 \citep{Alonso+08aa2}, HD\,189733 \citep{Pont+07aa2}, HD\,209458 \citep{Me09mn}, TrES-3 \citep{Sozzetti+09apj} and WASP-10 \citep{Johnson+09apj}. Error budgets indicate that a more precise velocity amplitude for the parent star would be useful for better measuring the physical properties of WASP-4\,b, but further improvement will require tackling the systematic errors which are unavoidable in the analysis of TEPs. The predictions of different sets of theoretical stellar models simply disagree for low mass stars, both between themselves and with the observed properties of low-mass eclipsing binary star systems. In the case of WASP-4, this disagreement is at the 5\% level for the mass of the planet and the 8\% level for the star's mass, with other properties less affected. This level of disagreement is a problem for inferring the chemical compositions of gas giant planets from their masses and radii \citep[see][]{Fortney++07apj}. For the best-studied cases, our understanding of TEPs therefore remains limited by our lack of understanding of low-mass stars.


\section*{Acknowledgments}

The reduced light curves presented in this work will be made available at the CDS ({\tt http://cdsweb.u-strasbg.fr/}) and at {\tt http://www.astro.keele.ac.uk/$\sim$jkt/}. We would like to thank the referee, John Johnson, for a positive and remarkably prompt report. JS acknowledges financial support from STFC in the form of a postdoctoral research position under the grant number ST/F002599/1. Astronomical research at the Armagh Observatory is funded by the Northern Ireland Department of Culture, Arts and Leisure (DCAL). DR (boursier FRIA) and J\,Surdej acknowledge support from the Communaut\'e fran\c{c}aise de Belgique - Actions de recherche concert\'ees - Acad\'emie Wallonie-Europe. The following internet-based resources were used in research for this paper: the ESO Digitized Sky Survey; the NASA Astrophysics Data System; the SIMBAD database operated at CDS, Strasbourg, France; and the ar$\chi$iv scientific paper preprint service operated by Cornell University.

\bibliographystyle{mn_new}


\appendix

\section{Results of the light curve analyses}

The tables in this section contain the full results of modelling each light curve of WASP-4.

\begin{table*} \caption{\label{tab:lceuler} Parameters of the {\sc jktebop} best
fits of the Euler $R$-band light curve of WASP-4 \citep{Wilson+08apj}, using
different approaches to LD. For each part of the table the upper quantities
are fitted parameters and the lower quantities are derived parameters. $T_0$
is given as HJD $-$ 2454000.0. The light curve contains 213 datapoints.}
\begin{tabular}{l r@{\,$\pm$\,}l r@{\,$\pm$\,}l r@{\,$\pm$\,}l r@{\,$\pm$\,}l r@{\,$\pm$\,}l}
\hline \hline
\                     & \mc{Linear LD law}                           & \mc{Quadratic LD law}                        & \mc{Square-root LD law}                      & \mc{Logarithmic LD law}                      & \mc{Cubic LD law}                            \\
\hline
\multicolumn{11}{l}{All LD coefficients fixed} \\
\hline
$r_{\rm A}+r_{\rm b}$ & \erm{0.2157}{0.0055}{0.0028}                 & \erm{0.2173}{0.0050}{0.0042}                 & \erm{0.2150}{0.0050}{0.0030}                 & \erm{0.2172}{0.0053}{0.0042}                 & \erm{0.2120}{0.0044}{0.0017}                 \\
$k$                   & \erm{0.15511}{0.00126}{0.00077}              & \erm{0.15631}{0.00094}{0.00079}              & \erm{0.15578}{0.00106}{0.00074}              & \erm{0.15605}{0.00108}{0.00084}              & \erm{0.15587}{0.00083}{0.00060}              \\
$i$ (deg.)            & \erm{88.15}{ 1.67}{ 1.27}                    & \erm{87.64}{ 1.63}{ 0.99}                    & \erm{88.21}{ 1.70}{ 1.19}                    & \erm{87.70}{ 1.80}{ 1.05}                    & \erm{89.15}{ 1.17}{ 1.31}                    \\
$u_{\rm A}$           & \mc{ 0.60 fixed}                             & \mc{ 0.40 fixed}                             & \mc{ 0.25 fixed}                             & \mc{ 0.70 fixed}                             & \mc{ 0.50 fixed}                             \\
$v_{\rm A}$           & \mc{ }                                       & \mc{ 0.25 fixed}                             & \mc{ 0.50 fixed}                             & \mc{ 0.23 fixed}                             & \mc{ 0.10 fixed}                             \\
$T_0$                 & \erm{365.91597}{  0.00011}{  0.00012}        & \erm{365.91599}{  0.00011}{  0.00010}        & \erm{365.91598}{  0.00011}{  0.00011}        & \erm{365.91598}{  0.00012}{  0.00011}        & \erm{365.91597}{  0.00011}{  0.00011}        \\
\hline
$r_{\rm A}$           & \erm{0.1867}{0.0045}{0.0023}                 & \erm{0.1879}{0.0041}{0.0035}                 & \erm{0.1860}{0.0042}{0.0025}                 & \erm{0.1878}{0.0044}{0.0035}                 & \erm{0.1834}{0.0036}{0.0014}                 \\
$r_{\rm b}$           & \erm{0.02897}{0.00095}{0.00046}              & \erm{0.02937}{0.00081}{0.00069}              & \erm{0.02898}{0.00086}{0.00048}              & \erm{0.02931}{0.00086}{0.00072}              & \erm{0.02859}{0.00075}{0.00027}              \\
$\sigma$ ($m$mag)     & \mc{ 1.9744}                                 & \mc{ 1.9437}                                 & \mc{ 1.9478}                                 & \mc{ 1.9483}                                 & \mc{ 1.9416}                                 \\
$\chi^2_{\rm \ red}$  & \mc{ 3.4310}                                 & \mc{ 3.3099}                                 & \mc{ 3.3258}                                 & \mc{ 3.3277}                                 & \mc{ 3.3011}                                 \\
\hline
\multicolumn{11}{l}{Fitting for the linear LD coefficient and fixing the nonlinear LD coefficient} \\
\hline
$r_{\rm A}+r_{\rm b}$ & \erm{0.2101}{0.0047}{0.0020}                 & \erm{0.2098}{0.0044}{0.0016}                 & \erm{0.2102}{0.0044}{0.0018}                 & \erm{0.2101}{0.0038}{0.0017}                 & \erm{0.2102}{0.0045}{0.0017}                 \\
$k$                   & \erm{0.15896}{0.00105}{0.00093}              & \erm{0.15760}{0.00119}{0.00091}              & \erm{0.15821}{0.00121}{0.00094}              & \erm{0.15792}{0.00109}{0.00096}              & \erm{0.15843}{0.00098}{0.00100}              \\
$i$ (deg.)            & \erm{88.67}{ 1.34}{ 1.25}                    & \erm{89.96}{ 0.94}{ 1.56}                    & \erm{88.94}{ 1.15}{ 1.37}                    & \erm{89.21}{ 1.11}{ 1.28}                    & \erm{88.84}{ 1.11}{ 1.38}                    \\
$u_{\rm A}$           & \erm{0.400}{0.032}{0.032}                    & \erm{0.296}{0.036}{0.039}                    & \erm{0.115}{0.037}{0.038}                    & \erm{0.572}{0.035}{0.040}                    & \erm{0.381}{0.035}{0.032}                    \\
$v_{\rm A}$           & \mc{ }                                       & \mc{ 0.25 fixed}                             & \mc{ 0.50 fixed}                             & \mc{ 0.23 fixed}                             & \mc{ 0.10 fixed}                             \\
$T_0$                 & \erm{365.91597}{  0.00012}{  0.00010}        & \erm{365.91598}{  0.00010}{  0.00011}        & \erm{365.91598}{  0.00011}{  0.00012}        & \erm{365.91598}{  0.00010}{  0.00011}        & \erm{365.91597}{  0.00011}{  0.00011}        \\
\hline
$r_{\rm A}$           & \erm{0.1812}{0.0040}{0.0017}                 & \erm{0.1812}{0.0037}{0.0015}                 & \erm{0.1815}{0.0036}{0.0016}                 & \erm{0.1814}{0.0031}{0.0015}                 & \erm{0.1815}{0.0037}{0.0016}                 \\
$r_{\rm b}$           & \erm{0.02881}{0.00081}{0.00030}              & \erm{0.02856}{0.00080}{0.00022}              & \erm{0.02872}{0.00077}{0.00026}              & \erm{0.02865}{0.00070}{0.00025}              & \erm{0.02875}{0.00078}{0.00026}              \\
$\sigma$ ($m$mag)     & \mc{ 1.9255}                                 & \mc{ 1.9261}                                 & \mc{ 1.9254}                                 & \mc{ 1.9255}                                 & \mc{ 1.9253}                                 \\
$\chi^2_{\rm \ red}$  & \mc{ 3.2528}                                 & \mc{ 3.2553}                                 & \mc{ 3.2527}                                 & \mc{ 3.2533}                                 & \mc{ 3.2523}                                 \\
\hline
\multicolumn{11}{l}{Fitting for the linear LD coefficient and perturbing the nonlinear LD coefficient} \\
\hline
$r_{\rm A}+r_{\rm b}$ & \mc{ }                                       & \erm{0.2098}{0.0038}{0.0015}                 & \erm{0.2102}{0.0043}{0.0018}                 & \erm{0.2101}{0.0042}{0.0017}                 & \erm{0.2102}{0.0048}{0.0019}                 \\
$k$                   & \mc{ }                                       & \erm{0.15760}{0.00111}{0.00104}              & \erm{0.15821}{0.00103}{0.00097}              & \erm{0.15792}{0.00112}{0.00099}              & \erm{0.15843}{0.00120}{0.00108}              \\
$i$ (deg.)            & \mc{ }                                       & \erm{89.96}{ 0.73}{ 1.56}                    & \erm{88.94}{ 1.28}{ 1.24}                    & \erm{89.21}{ 1.02}{ 1.48}                    & \erm{88.84}{ 1.27}{ 1.34}                    \\
$u_{\rm A}$           & \mc{ }                                       & \erm{0.296}{0.042}{0.045}                    & \erm{0.115}{0.049}{0.046}                    & \erm{0.572}{0.061}{0.059}                    & \erm{0.381}{0.034}{0.038}                    \\
$v_{\rm A}$           & \mc{ }                                       & \mc{ 0.25 perturbed}                         & \mc{ 0.50 perturbed}                         & \mc{ 0.23 perturbed}                         & \mc{ 0.10 perturbed}                         \\
$T_0$                 & \mc{ }                                       & \erm{365.915979}{  0.000095}{  0.000111}     & \erm{365.915975}{  0.000110}{  0.000103}     & \erm{365.915977}{  0.000101}{  0.000104}     & \erm{365.915975}{  0.000102}{  0.000107}     \\
\hline
$r_{\rm A}$           & \mc{ }                                       & \erm{0.1812}{0.0032}{0.0014}                 & \erm{0.1815}{0.0036}{0.0016}                 & \erm{0.1814}{0.0034}{0.0015}                 & \erm{0.1815}{0.0040}{0.0017}                 \\
$r_{\rm b}$           & \mc{ }                                       & \erm{0.02856}{0.00067}{0.00021}              & \erm{0.02872}{0.00076}{0.00026}              & \erm{0.02865}{0.00078}{0.00024}              & \erm{0.02875}{0.00082}{0.00030}              \\
$\sigma$ ($m$mag)     & \mc{ }                                       & \mc{ 1.9261}                                 & \mc{ 1.9254}                                 & \mc{ 1.9255}                                 & \mc{ 1.9253}                                 \\
$\chi^2_{\rm \ red}$  & \mc{ }                                       & \mc{ 3.2553}                                 & \mc{ 3.2527}                                 & \mc{ 3.2533}                                 & \mc{ 3.2523}                                 \\
\hline
\multicolumn{11}{l}{Fitting for both LD coefficients} \\
\hline
$r_{\rm A}+r_{\rm b}$ & \mc{ }                                       & \erm{0.2100}{0.0046}{0.0025}                 & \erm{0.2102}{0.0040}{0.0029}                 & \erm{0.2101}{0.0039}{0.0026}                 & \erm{0.2102}{0.0049}{0.0023}                 \\
$k$                   & \mc{ }                                       & \erm{0.15862}{0.00178}{0.00163}              & \erm{0.15860}{0.00215}{0.00227}              & \erm{0.15855}{0.00196}{0.00191}              & \erm{0.15849}{0.00202}{0.00197}              \\
$i$ (deg.)            & \mc{ }                                       & \erm{88.8}{ 1.3}{ 1.4}                       & \erm{88.8}{ 1.3}{ 1.5}                       & \erm{88.8}{ 1.3}{ 1.5}                       & \erm{88.8}{ 1.2}{ 1.6}                       \\
$u_{\rm A}$           & \mc{ }                                       & \erm{0.370}{0.119}{0.111}                    & \erm{0.252}{0.715}{0.698}                    & \erm{0.471}{0.269}{0.267}                    & \erm{0.382}{0.066}{0.068}                    \\
$v_{\rm A}$           & \mc{ }                                       & \erm{0.069}{0.253}{0.253}                    & \erm{0.259}{1.243}{1.243}                    & \erm{0.097}{0.336}{0.358}                    & \erm{0.090}{0.281}{0.295}                    \\
$T_0$                 & \mc{ }                                       & \erm{365.91597}{  0.00011}{  0.00012}        & \erm{365.91597}{  0.00011}{  0.00011}        & \erm{365.91597}{  0.00011}{  0.00011}        & \erm{365.91597}{  0.00011}{  0.00011}        \\
\hline
$r_{\rm A}$           & \mc{ }                                       & \erm{0.1813}{0.0036}{0.0022}                 & \erm{0.1815}{0.0032}{0.0027}                 & \erm{0.1814}{0.0033}{0.0022}                 & \erm{0.1815}{0.0039}{0.0020}                 \\
$r_{\rm b}$           & \mc{ }                                       & \erm{0.02876}{0.00093}{0.00031}              & \erm{0.02878}{0.00095}{0.00029}              & \erm{0.02875}{0.00091}{0.00030}              & \erm{0.02876}{0.00090}{0.00027}              \\
$\sigma$ ($m$mag)     & \mc{ }                                       & \mc{ 1.9253}                                 & \mc{ 1.9254}                                 & \mc{ 1.9253}                                 & \mc{ 1.9253}                                 \\
$\chi^2_{\rm \ red}$  & \mc{ }                                       & \mc{ 3.2682}                                 & \mc{ 3.2683}                                 & \mc{ 3.2682}                                 & \mc{ 3.2681}                                 \\
\hline \hline \end{tabular} \end{table*}

\begin{table*} \caption{\label{tab:lcvlt} Parameters of the {\sc jktebop}
best fits of the $z$-band VLT light curve of WASP-4 \citep{Gillon+09aa}, using
different approaches to LD. For each part of the table the upper quantities
are fitted parameters and the lower quantities are derived parameters.
$T_0$ is given as HJD $-$ 2454000.0. The light curves contain 244 datapoints.}
\begin{tabular}{l r@{\,$\pm$\,}l r@{\,$\pm$\,}l r@{\,$\pm$\,}l r@{\,$\pm$\,}l r@{\,$\pm$\,}l}
\hline \hline
\                     & \mc{Linear LD law}                           & \mc{Quadratic LD law}                        & \mc{Square-root LD law}                      & \mc{Logarithmic LD law}                      & \mc{Cubic LD law}                            \\
\hline
\multicolumn{11}{l}{All LD coefficients fixed} \\
\hline
$r_{\rm A}+r_{\rm b}$ & \erm{0.2085}{0.0017}{0.0004}                 & \erm{0.2099}{0.0015}{0.0014}                 & \erm{0.2094}{0.0016}{0.0015}                 & \erm{0.2094}{0.0015}{0.0015}                 & \erm{0.2092}{0.0016}{0.0015}                 \\
$k$                   & \erm{0.15167}{0.00035}{0.00023}              & \erm{0.15304}{0.00025}{0.00025}              & \erm{0.15313}{0.00027}{0.00026}              & \erm{0.15308}{0.00027}{0.00025}              & \erm{0.15313}{0.00031}{0.00028}              \\
$i$ (deg.)            & \erm{89.91}{ 0.46}{ 0.95}                    & \erm{88.56}{ 0.70}{ 0.47}                    & \erm{88.61}{ 0.82}{ 0.51}                    & \erm{88.64}{ 0.84}{ 0.53}                    & \erm{88.65}{ 0.95}{ 0.51}                    \\
$u_{\rm A}$           & \mc{ 0.50 fixed}                             & \mc{ 0.25 fixed}                             & \mc{ 0.10 fixed}                             & \mc{ 0.59 fixed}                             & \mc{ 0.40 fixed}                             \\
$v_{\rm A}$           & \mc{ }                                       & \mc{ 0.31 fixed}                             & \mc{ 0.54 fixed}                             & \mc{ 0.26 fixed}                             & \mc{ 0.10 fixed}                             \\
$T_0$                 & \erm{396.695401}{  0.000037}{  0.000036}     & \erm{396.695391}{  0.000035}{  0.000035}     & \erm{396.695394}{  0.000036}{  0.000035}     & \erm{396.695393}{  0.000036}{  0.000033}     & \erm{396.695395}{  0.000035}{  0.000037}     \\
\hline
$r_{\rm A}$           & \erm{0.1810}{0.0014}{0.0004}                 & \erm{0.1820}{0.0013}{0.0012}                 & \erm{0.1816}{0.0013}{0.0013}                 & \erm{0.1816}{0.0013}{0.0013}                 & \erm{0.1814}{0.0013}{0.0013}                 \\
$r_{\rm b}$           & \erm{0.027451}{0.000271}{0.000076}           & \erm{0.02785}{0.00023}{0.00023}              & \erm{0.02780}{0.00026}{0.00023}              & \erm{0.02781}{0.00024}{0.00023}              & \erm{0.02778}{0.00025}{0.00024}              \\
$\sigma$ ($m$mag)     & \mc{ 0.6650}                                 & \mc{ 0.6111}                                 & \mc{ 0.6056}                                 & \mc{ 0.6059}                                 & \mc{ 0.6075}                                 \\
$\chi^2_{\rm \ red}$  & \mc{ 2.0671}                                 & \mc{ 1.6927}                                 & \mc{ 1.6552}                                 & \mc{ 1.6571}                                 & \mc{ 1.6680}                                 \\
\hline
\multicolumn{11}{l}{Fitting for the linear LD coefficient and fixing the nonlinear LD coefficient} \\
\hline
$r_{\rm A}+r_{\rm b}$ & \erm{0.2103}{0.0016}{0.0016}                 & \erm{0.2086}{0.0017}{0.0007}                 & \erm{0.2099}{0.0015}{0.0017}                 & \erm{0.2094}{0.0017}{0.0014}                 & \erm{0.2099}{0.0016}{0.0018}                 \\
$k$                   & \erm{0.15442}{0.00041}{0.00040}              & \erm{0.15244}{0.00044}{0.00033}              & \erm{0.15351}{0.00042}{0.00039}              & \erm{0.15307}{0.00045}{0.00037}              & \erm{0.15378}{0.00044}{0.00043}              \\
$i$ (deg.)            & \erm{87.96}{ 0.49}{ 0.39}                    & \erm{89.42}{ 0.66}{ 0.90}                    & \erm{88.33}{ 0.63}{ 0.47}                    & \erm{88.65}{ 0.95}{ 0.59}                    & \erm{88.25}{ 0.68}{ 0.44}                    \\
$u_{\rm A}$           & \erm{0.392}{0.010}{0.011}                    & \erm{0.270}{0.013}{0.013}                    & \erm{0.086}{0.011}{0.011}                    & \erm{0.590}{0.012}{0.011}                    & \erm{0.375}{0.012}{0.012}                    \\
$v_{\rm A}$           & \mc{ }                                       & \mc{ 0.31 fixed}                             & \mc{ 0.54 fixed}                             & \mc{ 0.26 fixed}                             & \mc{ 0.10 fixed}                             \\
$T_0$                 & \erm{396.695394}{  0.000034}{  0.000035}     & \erm{396.695393}{  0.000035}{  0.000034}     & \erm{396.695393}{  0.000038}{  0.000036}     & \erm{396.695393}{  0.000034}{  0.000037}     & \erm{396.695393}{  0.000035}{  0.000036}     \\
\hline
$r_{\rm A}$           & \erm{0.1821}{0.0013}{0.0013}                 & \erm{0.1810}{0.0014}{0.0006}                 & \erm{0.1820}{0.0013}{0.0014}                 & \erm{0.1816}{0.0014}{0.0012}                 & \erm{0.1819}{0.0014}{0.0015}                 \\
$r_{\rm b}$           & \erm{0.02813}{0.00026}{0.00026}              & \erm{0.02759}{0.00028}{0.00010}              & \erm{0.02793}{0.00027}{0.00027}              & \erm{0.02780}{0.00029}{0.00023}              & \erm{0.02797}{0.00027}{0.00029}              \\
$\sigma$ ($m$mag)     & \mc{ 0.6046}                                 & \mc{ 0.6095}                                 & \mc{ 0.6049}                                 & \mc{ 0.6059}                                 & \mc{ 0.6046}                                 \\
$\chi^2_{\rm \ red}$  & \mc{ 1.6533}                                 & \mc{ 1.6900}                                 & \mc{ 1.6566}                                 & \mc{ 1.6640}                                 & \mc{ 1.6538}                                 \\
\hline
\multicolumn{11}{l}{Fitting for the linear LD coefficient and perturbing the nonlinear LD coefficient} \\
\hline
$r_{\rm A}+r_{\rm b}$ & \mc{ }                                       & \erm{0.2086}{0.0016}{0.0008}                 & \erm{0.2099}{0.0016}{0.0016}                 & \erm{0.2094}{0.0018}{0.0013}                 & \erm{0.2099}{0.0015}{0.0017}                 \\
$k$                   & \mc{ }                                       & \erm{0.15244}{0.00055}{0.00048}              & \erm{0.15351}{0.00045}{0.00039}              & \erm{0.15307}{0.00057}{0.00049}              & \erm{0.15378}{0.00058}{0.00060}              \\
$i$ (deg.)            & \mc{ }                                       & \erm{89.42}{ 0.71}{ 0.84}                    & \erm{88.33}{ 0.65}{ 0.47}                    & \erm{88.65}{ 1.02}{ 0.62}                    & \erm{88.25}{ 0.68}{ 0.45}                    \\
$u_{\rm A}$           & \mc{ }                                       & \erm{0.270}{0.027}{0.029}                    & \erm{0.086}{0.038}{0.039}                    & \erm{0.590}{0.051}{0.053}                    & \erm{0.375}{0.016}{0.016}                    \\
$v_{\rm A}$           & \mc{ }                                       & \mc{ 0.31 perturbed}                         & \mc{ 0.54 perturbed}                         & \mc{ 0.26 perturbed}                         & \mc{ 0.10 perturbed}                         \\
$T_0$                 & \mc{ }                                       & \erm{396.695393}{  0.000037}{  0.000036}     & \erm{396.695393}{  0.000033}{  0.000034}     & \erm{396.695393}{  0.000038}{  0.000032}     & \erm{396.695393}{  0.000034}{  0.000037}     \\
\hline
$r_{\rm A}$           & \mc{ }                                       & \erm{0.1810}{0.0013}{0.0007}                 & \erm{0.1820}{0.0013}{0.0014}                 & \erm{0.1816}{0.0015}{0.0011}                 & \erm{0.1819}{0.0013}{0.0014}                 \\
$r_{\rm b}$           & \mc{ }                                       & \erm{0.02759}{0.00030}{0.00010}              & \erm{0.02793}{0.00027}{0.00028}              & \erm{0.02780}{0.00029}{0.00024}              & \erm{0.02797}{0.00026}{0.00030}              \\
$\sigma$ ($m$mag)     & \mc{ }                                       & \mc{ 0.6095}                                 & \mc{ 0.6049}                                 & \mc{ 0.6059}                                 & \mc{ 0.6046}                                 \\
$\chi^2_{\rm \ red}$  & \mc{ }                                       & \mc{ 1.6900}                                 & \mc{ 1.6566}                                 & \mc{ 1.6640}                                 & \mc{ 1.6538}                                 \\
\hline
\multicolumn{11}{l}{Fitting for both LD coefficients} \\
\hline
$r_{\rm A}+r_{\rm b}$ & \mc{ }                                       & \erm{0.2100}{0.0017}{0.0017}                 & \erm{0.2101}{0.0016}{0.0016}                 & \erm{0.2100}{0.0017}{0.0015}                 & \erm{0.2101}{0.0017}{0.0017}                 \\
$k$                   & \mc{ }                                       & \erm{0.15412}{0.00067}{0.00076}              & \erm{0.15412}{0.00080}{0.00089}              & \erm{0.15412}{0.00078}{0.00073}              & \erm{0.15417}{0.00076}{0.00087}              \\
$i$ (deg.)            & \mc{ }                                       & \erm{88.12}{ 0.72}{ 0.51}                    & \erm{88.09}{ 0.76}{ 0.49}                    & \erm{88.11}{ 0.69}{ 0.49}                    & \erm{88.08}{ 0.74}{ 0.48}                    \\
$u_{\rm A}$           & \mc{ }                                       & \erm{0.371}{0.037}{0.039}                    & \erm{0.290}{0.250}{0.248}                    & \erm{0.437}{0.091}{0.097}                    & \erm{0.385}{0.023}{0.023}                    \\
$v_{\rm A}$           & \mc{ }                                       & \erm{0.050}{0.091}{0.083}                    & \erm{0.179}{0.425}{0.440}                    & \erm{0.059}{0.125}{0.123}                    & \erm{0.039}{0.115}{0.099}                    \\
$T_0$                 & \mc{ }                                       & \erm{396.695393}{  0.000034}{  0.000035}     & \erm{396.695393}{  0.000033}{  0.000035}     & \erm{396.695393}{  0.000036}{  0.000034}     & \erm{396.695393}{  0.000036}{  0.000038}     \\
\hline
$r_{\rm A}$           & \mc{ }                                       & \erm{0.1819}{0.0014}{0.0014}                 & \erm{0.1820}{0.0013}{0.0013}                 & \erm{0.1820}{0.0014}{0.0013}                 & \erm{0.1820}{0.0014}{0.0014}                 \\
$r_{\rm b}$           & \mc{ }                                       & \erm{0.02804}{0.00032}{0.00032}              & \erm{0.02805}{0.00031}{0.00035}              & \erm{0.02805}{0.00031}{0.00030}              & \erm{0.02806}{0.00032}{0.00033}              \\
$\sigma$ ($m$mag)     & \mc{ }                                       & \mc{ 0.6042}                                 & \mc{ 0.6044}                                 & \mc{ 0.6044}                                 & \mc{ 0.6044}                                 \\
$\chi^2_{\rm \ red}$  & \mc{ }                                       & \mc{ 1.6586}                                 & \mc{ 1.6596}                                 & \mc{ 1.6592}                                 & \mc{ 1.6595}                                 \\
\hline \hline \end{tabular} \end{table*}

\begin{table*} \caption{\label{tab:lcmag} Parameters of the {\sc jktebop} best
fits of the Magellan $z$-band telescope light curve of WASP-4 \citep{Winn+09aj},
using different approaches to LD. For each part of the table the upper quantities
are fitted parameters and the lower quantities are derived parameters. $T_0$ is
given as HJD $-$ 2454000.0. The light curves contain 713 datapoints.}
\begin{tabular}{l r@{\,$\pm$\,}l r@{\,$\pm$\,}l r@{\,$\pm$\,}l r@{\,$\pm$\,}l r@{\,$\pm$\,}l}
\hline \hline
\                     & \mc{Linear LD law}                           & \mc{Quadratic LD law}                        & \mc{Square-root LD law}                      & \mc{Logarithmic LD law}                      & \mc{Cubic LD law}                            \\
\hline
\multicolumn{11}{l}{All LD coefficients fixed} \\
\hline
$r_{\rm A}+r_{\rm b}$ & \erm{0.2103}{0.0014}{0.0004}                 & \erm{0.2126}{0.0014}{0.0014}                 & \erm{0.2122}{0.0015}{0.0013}                 & \erm{0.2124}{0.0014}{0.0016}                 & \erm{0.2119}{0.0015}{0.0013}                 \\
$k$                   & \erm{0.15299}{0.00028}{0.00021}              & \erm{0.15405}{0.00026}{0.00025}              & \erm{0.15428}{0.00030}{0.00025}              & \erm{0.15419}{0.00026}{0.00029}              & \erm{0.15431}{0.00028}{0.00030}              \\
$i$ (deg.)            & \erm{89.63}{ 0.56}{ 0.79}                    & \erm{88.26}{ 0.51}{ 0.37}                    & \erm{88.25}{ 0.45}{ 0.40}                    & \erm{88.24}{ 0.51}{ 0.37}                    & \erm{88.31}{ 0.50}{ 0.40}                    \\
$u_{\rm A}$           & \mc{ 0.50 fixed}                             & \mc{ 0.25 fixed}                             & \mc{ 0.10 fixed}                             & \mc{ 0.59 fixed}                             & \mc{ 0.40 fixed}                             \\
$v_{\rm A}$           & \mc{ }                                       & \mc{ 0.31 fixed}                             & \mc{ 0.54 fixed}                             & \mc{ 0.26 fixed}                             & \mc{ 0.10 fixed}                             \\
$T_0$                 & \erm{697.797572}{  0.000032}{  0.000038}     & \erm{697.797565}{  0.000034}{  0.000032}     & \erm{697.797567}{  0.000032}{  0.000036}     & \erm{697.797566}{  0.000033}{  0.000035}     & \erm{697.797568}{  0.000036}{  0.000032}     \\
\hline
$r_{\rm A}$           & \erm{0.1824}{0.0012}{0.0004}                 & \erm{0.1842}{0.0012}{0.0012}                 & \erm{0.1839}{0.0012}{0.0011}                 & \erm{0.1840}{0.0012}{0.0013}                 & \erm{0.1836}{0.0012}{0.0011}                 \\
$r_{\rm b}$           & \erm{0.02790}{0.00024}{0.00008}              & \erm{0.02838}{0.00023}{0.00022}              & \erm{0.02837}{0.00024}{0.00021}              & \erm{0.02838}{0.00021}{0.00024}              & \erm{0.02833}{0.00023}{0.00022}              \\
$\sigma$ ($m$mag)     & \mc{ 0.6479}                                 & \mc{ 0.6039}                                 & \mc{ 0.6013}                                 & \mc{ 0.6012}                                 & \mc{ 0.6027}                                 \\
$\chi^2_{\rm \ red}$  & \mc{ 0.7892}                                 & \mc{ 0.7003}                                 & \mc{ 0.6944}                                 & \mc{ 0.6947}                                 & \mc{ 0.6964}                                 \\
\hline
\multicolumn{11}{l}{Fitting for the linear LD coefficient and fixing the nonlinear LD coefficient} \\
\hline
$r_{\rm A}+r_{\rm b}$ & \erm{0.2123}{0.0015}{0.0014}                 & \erm{0.2112}{0.0016}{0.0009}                 & \erm{0.2122}{0.0016}{0.0016}                 & \erm{0.2118}{0.0015}{0.0014}                 & \erm{0.2122}{0.0014}{0.0015}                 \\
$k$                   & \erm{0.15506}{0.00039}{0.00039}              & \erm{0.15336}{0.00046}{0.00033}              & \erm{0.15424}{0.00041}{0.00046}              & \erm{0.15387}{0.00044}{0.00041}              & \erm{0.15452}{0.00042}{0.00041}              \\
$i$ (deg.)            & \erm{88.00}{ 0.44}{ 0.39}                    & \erm{88.98}{ 0.92}{ 0.68}                    & \erm{88.27}{ 0.61}{ 0.46}                    & \erm{88.51}{ 0.70}{ 0.50}                    & \erm{88.18}{ 0.53}{ 0.40}                    \\
$u_{\rm A}$           & \erm{0.412}{0.011}{0.012}                    & \erm{0.276}{0.012}{0.014}                    & \erm{0.102}{0.012}{0.013}                    & \erm{0.603}{0.012}{0.013}                    & \erm{0.391}{0.011}{0.012}                    \\
$v_{\rm A}$           & \mc{ }                                       & \mc{ 0.31 fixed}                             & \mc{ 0.54 fixed}                             & \mc{ 0.26 fixed}                             & \mc{ 0.10 fixed}                             \\
$T_0$                 & \erm{697.797568}{  0.000032}{  0.000035}     & \erm{697.797565}{  0.000036}{  0.000033}     & \erm{697.797567}{  0.000034}{  0.000034}     & \erm{697.797566}{  0.000034}{  0.000032}     & \erm{697.797567}{  0.000034}{  0.000034}     \\
\hline
$r_{\rm A}$           & \erm{0.1838}{0.0012}{0.0012}                 & \erm{0.1831}{0.0013}{0.0007}                 & \erm{0.1838}{0.0013}{0.0013}                 & \erm{0.1836}{0.0013}{0.0012}                 & \erm{0.1838}{0.0012}{0.0013}                 \\
$r_{\rm b}$           & \erm{0.02849}{0.00025}{0.00023}              & \erm{0.02808}{0.00027}{0.00015}              & \erm{0.02836}{0.00027}{0.00027}              & \erm{0.02825}{0.00026}{0.00024}              & \erm{0.02840}{0.00025}{0.00025}              \\
$\sigma$ ($m$mag)     & \mc{ 0.6039}                                 & \mc{ 0.6020}                                 & \mc{ 0.6014}                                 & \mc{ 0.6009}                                 & \mc{ 0.6018}                                 \\
$\chi^2_{\rm \ red}$  & \mc{ 0.7014}                                 & \mc{ 0.6961}                                 & \mc{ 0.6954}                                 & \mc{ 0.6943}                                 & \mc{ 0.6965}                                 \\
\hline
\multicolumn{11}{l}{Fitting for the linear LD coefficient and perturbing the nonlinear LD coefficient} \\
\hline
$r_{\rm A}+r_{\rm b}$ & \mc{ }                                       & \erm{0.2112}{0.0016}{0.0009}                 & \erm{0.2122}{0.0015}{0.0016}                 & \erm{0.2118}{0.0016}{0.0015}                 & \erm{0.2122}{0.0015}{0.0016}                 \\
$k$                   & \mc{ }                                       & \erm{0.15336}{0.00060}{0.00049}              & \erm{0.15424}{0.00047}{0.00040}              & \erm{0.15387}{0.00053}{0.00054}              & \erm{0.15452}{0.00058}{0.00059}              \\
$i$ (deg.)            & \mc{ }                                       & \erm{88.98}{ 0.91}{ 0.77}                    & \erm{88.27}{ 0.62}{ 0.45}                    & \erm{88.51}{ 0.90}{ 0.50}                    & \erm{88.18}{ 0.62}{ 0.43}                    \\
$u_{\rm A}$           & \mc{ }                                       & \erm{0.276}{0.030}{0.029}                    & \erm{0.102}{0.038}{0.039}                    & \erm{0.603}{0.054}{0.052}                    & \erm{0.391}{0.016}{0.018}                    \\
$v_{\rm A}$           & \mc{ }                                       & \mc{ 0.31 perturbed}                         & \mc{ 0.54 perturbed}                         & \mc{ 0.26 perturbed}                         & \mc{ 0.10 perturbed}                         \\
$T_0$                 & \mc{ }                                       & \erm{697.797565}{  0.000033}{  0.000036}     & \erm{697.797567}{  0.000033}{  0.000032}     & \erm{697.797566}{  0.000033}{  0.000033}     & \erm{697.797567}{  0.000032}{  0.000034}     \\
\hline
$r_{\rm A}$           & \mc{ }                                       & \erm{0.1831}{0.0014}{0.0008}                 & \erm{0.1838}{0.0012}{0.0013}                 & \erm{0.1836}{0.0014}{0.0012}                 & \erm{0.1838}{0.0013}{0.0014}                 \\
$r_{\rm b}$           & \mc{ }                                       & \erm{0.02808}{0.00031}{0.00015}              & \erm{0.02836}{0.00027}{0.00026}              & \erm{0.02825}{0.00027}{0.00029}              & \erm{0.02840}{0.00025}{0.00029}              \\
$\sigma$ ($m$mag)     & \mc{ }                                       & \mc{ 0.6020}                                 & \mc{ 0.6014}                                 & \mc{ 0.6009}                                 & \mc{ 0.6018}                                 \\
$\chi^2_{\rm \ red}$  & \mc{ }                                       & \mc{ 0.6961}                                 & \mc{ 0.6954}                                 & \mc{ 0.6943}                                 & \mc{ 0.6965}                                 \\
\hline
\multicolumn{11}{l}{Fitting for both LD coefficients} \\
\hline
$r_{\rm A}+r_{\rm b}$ & \mc{ }                                       & \erm{0.2116}{0.0016}{0.0014}                 & \erm{0.2119}{0.0017}{0.0012}                 & \erm{0.2118}{0.0017}{0.0013}                 & \erm{0.2118}{0.0016}{0.0013}                 \\
$k$                   & \mc{ }                                       & \erm{0.15400}{0.00071}{0.00066}              & \erm{0.15356}{0.00090}{0.00086}              & \erm{0.15376}{0.00089}{0.00077}              & \erm{0.15363}{0.00093}{0.00086}              \\
$i$ (deg.)            & \mc{ }                                       & \erm{88.52}{ 1.03}{ 0.59}                    & \erm{88.60}{ 1.23}{ 0.66}                    & \erm{88.57}{ 1.16}{ 0.65}                    & \erm{88.61}{ 1.16}{ 0.65}                    \\
$u_{\rm A}$           & \mc{ }                                       & \erm{0.32}{0.04}{0.04}                       & \erm{$-$0.12}{0.25}{0.26}                    & \erm{0.62}{0.10}{0.10}                       & \erm{0.37}{0.02}{0.02}                       \\
$v_{\rm A}$           & \mc{ }                                       & \erm{0.202}{0.098}{0.094}                    & \erm{0.928}{0.457}{0.438}                    & \erm{0.282}{0.137}{0.134}                    & \erm{0.240}{0.130}{0.123}                    \\
$T_0$                 & \mc{ }                                       & \erm{697.797566}{  0.000034}{  0.000034}     & \erm{697.797566}{  0.000035}{  0.000034}     & \erm{697.797566}{  0.000033}{  0.000034}     & \erm{697.797566}{  0.000035}{  0.000036}     \\
\hline
$r_{\rm A}$           & \mc{ }                                       & \erm{0.1834}{0.0013}{0.0011}                 & \erm{0.1837}{0.0014}{0.0010}                 & \erm{0.1835}{0.0014}{0.0011}                 & \erm{0.1836}{0.0012}{0.0011}                 \\
$r_{\rm b}$           & \mc{ }                                       & \erm{0.02824}{0.00031}{0.00026}              & \erm{0.02821}{0.00035}{0.00025}              & \erm{0.02822}{0.00033}{0.00027}              & \erm{0.02821}{0.00032}{0.00026}              \\
$\sigma$ ($m$mag)     & \mc{ }                                       & \mc{ 0.6009}                                 & \mc{ 0.6010}                                 & \mc{ 0.6009}                                 & \mc{ 0.6010}                                 \\
$\chi^2_{\rm \ red}$  & \mc{ }                                       & \mc{ 0.6952}                                 & \mc{ 0.6953}                                 & \mc{ 0.6952}                                 & \mc{ 0.6952}                                 \\
\hline \hline \end{tabular} \end{table*}

\begin{table*} \caption{\label{tab:lcdanish} Parameters of the {\sc jktebop}
best fits of the Danish telescope $R$-band light curve of WASP-4 (this work),
using different approaches to LD. For each part of the table the upper
quantities are fitted parameters and the lower quantities are derived
parameters. $T_0$ is given as HJD $-$ 2454000.0. The light curves contain 452 datapoints.}
\begin{tabular}{l r@{\,$\pm$\,}l r@{\,$\pm$\,}l r@{\,$\pm$\,}l r@{\,$\pm$\,}l r@{\,$\pm$\,}l}
\hline \hline
\                     & \mc{Linear LD law}                           & \mc{Quadratic LD law}                        & \mc{Square-root LD law}                      & \mc{Logarithmic LD law}                      & \mc{Cubic LD law}                            \\
\hline
\multicolumn{11}{l}{All LD coefficients fixed} \\
\hline
$r_{\rm A}+r_{\rm b}$ & \erm{0.2103}{0.0021}{0.0006}                 & \erm{0.2097}{0.0018}{0.0006}                 & \erm{0.2097}{0.0020}{0.0006}                 & \erm{0.2099}{0.0021}{0.0005}                 & \erm{0.2092}{0.0020}{0.0005}                 \\
$k$                   & \erm{0.15245}{0.00045}{0.00029}              & \erm{0.15345}{0.00033}{0.00023}              & \erm{0.15322}{0.00040}{0.00025}              & \erm{0.15316}{0.00039}{0.00024}              & \erm{0.15377}{0.00037}{0.00026}              \\
$i$ (deg.)            & \erm{89.84}{ 0.61}{ 1.04}                    & \erm{89.82}{ 0.59}{ 0.91}                    & \erm{89.97}{ 0.52}{ 1.04}                    & \erm{89.97}{ 0.40}{ 1.17}                    & \erm{89.73}{ 0.56}{ 0.97}                    \\
$u_{\rm A}$           & \mc{ 0.60 fixed}                             & \mc{ 0.40 fixed}                             & \mc{ 0.25 fixed}                             & \mc{ 0.70 fixed}                             & \mc{ 0.50 fixed}                             \\
$v_{\rm A}$           & \mc{ }                                       & \mc{ 0.25 fixed}                             & \mc{ 0.50 fixed}                             & \mc{ 0.23 fixed}                             & \mc{ 0.10 fixed}                             \\
$T_0$                 & \erm{365.916894}{  0.000047}{  0.000045}     & \erm{365.916893}{  0.000041}{  0.000047}     & \erm{365.916894}{  0.000041}{  0.000047}     & \erm{365.916894}{  0.000045}{  0.000049}     & \erm{365.916894}{  0.000043}{  0.000046}     \\
\hline
$r_{\rm A}$           & \erm{0.1825}{0.0018}{0.0005}                 & \erm{0.1818}{0.0015}{0.0005}                 & \erm{0.1819}{0.0017}{0.0005}                 & \erm{0.1820}{0.0018}{0.0004}                 & \erm{0.1813}{0.0016}{0.0004}                 \\
$r_{\rm b}$           & \erm{0.02782}{0.00035}{0.00011}              & \erm{0.02790}{0.00029}{0.00010}              & \erm{0.02786}{0.00034}{0.00009}              & \erm{0.02787}{0.00035}{0.00008}              & \erm{0.02788}{0.00032}{0.00009}              \\
$\sigma$ ($m$mag)     & \mc{ 0.8792}                                 & \mc{ 0.8576}                                 & \mc{ 0.8558}                                 & \mc{ 0.8555}                                 & \mc{ 0.8553}                                 \\
$\chi^2_{\rm \ red}$  & \mc{ 1.4308}                                 & \mc{ 1.3189}                                 & \mc{ 1.3187}                                 & \mc{ 1.3164}                                 & \mc{ 1.3107}                                 \\
\hline
\multicolumn{11}{l}{Fitting for the linear LD coefficient and fixing the nonlinear LD coefficient} \\
\hline
$r_{\rm A}+r_{\rm b}$ & \erm{0.2085}{0.0019}{0.0008}                 & \erm{0.2099}{0.0021}{0.0008}                 & \erm{0.2093}{0.0020}{0.0008}                 & \erm{0.2094}{0.0020}{0.0007}                 & \erm{0.2088}{0.0019}{0.0007}                 \\
$k$                   & \erm{0.15471}{0.00054}{0.00041}              & \erm{0.15344}{0.00066}{0.00045}              & \erm{0.15392}{0.00052}{0.00045}              & \erm{0.15363}{0.00054}{0.00045}              & \erm{0.15407}{0.00058}{0.00041}              \\
$i$ (deg.)            & \erm{89.30}{ 0.76}{ 0.86}                    & \erm{89.60}{ 0.65}{ 1.08}                    & \erm{89.50}{ 0.72}{ 0.95}                    & \erm{89.95}{ 0.63}{ 1.03}                    & \erm{90.00}{ 0.50}{ 1.12}                    \\
$u_{\rm A}$           & \erm{0.501}{0.013}{0.013}                    & \erm{0.401}{0.014}{0.017}                    & \erm{0.220}{0.014}{0.015}                    & \erm{0.678}{0.014}{0.015}                    & \erm{0.485}{0.015}{0.015}                    \\
$v_{\rm A}$           & \mc{ }                                       & \mc{ 0.25 fixed}                             & \mc{ 0.50 fixed}                             & \mc{ 0.23 fixed}                             & \mc{ 0.10 fixed}                             \\
$T_0$                 & \erm{365.916894}{  0.000039}{  0.000047}     & \erm{365.916893}{  0.000045}{  0.000044}     & \erm{365.916894}{  0.000044}{  0.000046}     & \erm{365.916893}{  0.000046}{  0.000046}     & \erm{365.916894}{  0.000047}{  0.000043}     \\
\hline
$r_{\rm A}$           & \erm{0.1806}{0.0015}{0.0007}                 & \erm{0.1819}{0.0017}{0.0007}                 & \erm{0.1814}{0.0016}{0.0007}                 & \erm{0.1815}{0.0016}{0.0007}                 & \erm{0.1809}{0.0016}{0.0006}                 \\
$r_{\rm b}$           & \erm{0.02794}{0.00033}{0.00013}              & \erm{0.02792}{0.00041}{0.00010}              & \erm{0.02791}{0.00037}{0.00011}              & \erm{0.02788}{0.00037}{0.00010}              & \erm{0.02787}{0.00036}{0.00008}              \\
$\sigma$ ($m$mag)     & \mc{ 0.8568}                                 & \mc{ 0.8575}                                 & \mc{ 0.8559}                                 & \mc{ 0.8562}                                 & \mc{ 0.8559}                                 \\
$\chi^2_{\rm \ red}$  & \mc{ 1.3115}                                 & \mc{ 1.3222}                                 & \mc{ 1.3116}                                 & \mc{ 1.3138}                                 & \mc{ 1.3109}                                 \\
\hline
\multicolumn{11}{l}{Fitting for the linear LD coefficient and perturbing the nonlinear LD coefficient} \\
\hline
$r_{\rm A}+r_{\rm b}$ & \mc{ }                                       & \erm{0.2099}{0.0012}{0.0009}                 & \erm{0.2093}{0.0020}{0.0008}                 & \erm{0.2094}{0.0020}{0.0008}                 & \erm{0.2088}{0.0018}{0.0008}                 \\
$k$                   & \mc{ }                                       & \erm{0.15344}{0.00069}{0.00064}              & \erm{0.15392}{0.00061}{0.00043}              & \erm{0.15363}{0.00064}{0.00052}              & \erm{0.15407}{0.00070}{0.00058}              \\
$i$ (deg.)            & \mc{ }                                       & \erm{89.60}{ 0.63}{ 1.08}                    & \erm{89.50}{ 0.58}{ 1.09}                    & \erm{89.95}{ 0.58}{ 1.06}                    & \erm{90.00}{ 0.52}{ 1.10}                    \\
$u_{\rm A}$           & \mc{ }                                       & \erm{0.401}{0.029}{0.028}                    & \erm{0.220}{0.038}{0.041}                    & \erm{0.678}{0.055}{0.051}                    & \erm{0.485}{0.017}{0.018}                    \\
$v_{\rm A}$           & \mc{ }                                       & \mc{ 0.25 perturbed}                         & \mc{ 0.50 perturbed}                         & \mc{ 0.23 perturbed}                         & \mc{ 0.10 perturbed}                         \\
$T_0$                 & \mc{ }                                       & \erm{365.916893}{  0.000045}{  0.000046}     & \erm{365.916894}{  0.000046}{  0.000043}     & \erm{365.916893}{  0.000047}{  0.000046}     & \erm{365.916894}{  0.000042}{  0.000045}     \\
\hline
$r_{\rm A}$           & \mc{ }                                       & \erm{0.1819}{0.0016}{0.0008}                 & \erm{0.1814}{0.0016}{0.0007}                 & \erm{0.1815}{0.0016}{0.0008}                 & \erm{0.1809}{0.0015}{0.0008}                 \\
$r_{\rm b}$           & \mc{ }                                       & \erm{0.02792}{0.00040}{0.00011}              & \erm{0.02791}{0.00037}{0.00010}              & \erm{0.02788}{0.00037}{0.00009}              & \erm{0.02787}{0.00036}{0.00009}              \\
$\sigma$ ($m$mag)     & \mc{ }                                       & \mc{ 0.8575}                                 & \mc{ 0.8559}                                 & \mc{ 0.8562}                                 & \mc{ 0.8559}                                 \\
$\chi^2_{\rm \ red}$  & \mc{ }                                       & \mc{ 1.3222}                                 & \mc{ 1.3116}                                 & \mc{ 1.3138}                                 & \mc{ 1.3109}                                 \\
\hline
\multicolumn{11}{l}{Fitting for both LD coefficients} \\
\hline
$r_{\rm A}+r_{\rm b}$ & \mc{ }                                       & \erm{0.2088}{0.0023}{0.0011}                 & \erm{0.2093}{0.0019}{0.0014}                 & \erm{0.2091}{0.0022}{0.0012}                 & \erm{0.2093}{0.0022}{0.0015}                 \\
$k$                   & \mc{ }                                       & \erm{0.15428}{0.00103}{0.00075}              & \erm{0.15409}{0.00116}{0.00103}              & \erm{0.15423}{0.00123}{0.00086}              & \erm{0.15426}{0.00130}{0.00106}              \\
$i$ (deg.)            & \mc{ }                                       & \erm{90.00}{ 0.60}{ 1.34}                    & \erm{89.58}{ 0.67}{ 1.23}                    & \erm{89.57}{ 0.75}{ 1.13}                    & \erm{89.36}{ 0.92}{ 1.09}                    \\
$u_{\rm A}$           & \mc{ }                                       & \erm{0.476}{0.046}{0.047}                    & \erm{0.276}{0.291}{0.294}                    & \erm{0.595}{0.117}{0.134}                    & \erm{0.493}{0.028}{0.028}                    \\
$v_{\rm A}$           & \mc{ }                                       & \erm{0.081}{0.108}{0.113}                    & \erm{0.412}{0.531}{0.516}                    & \erm{0.114}{0.151}{0.169}                    & \erm{0.093}{0.144}{0.155}                    \\
$T_0$                 & \mc{ }                                       & \erm{365.916885}{  0.000052}{  0.000047}     & \erm{365.916885}{  0.000051}{  0.000050}     & \erm{365.916885}{  0.000052}{  0.000052}     & \erm{365.916884}{  0.000050}{  0.000053}     \\
\hline
$r_{\rm A}$           & \mc{ }                                       & \erm{0.1809}{0.0019}{0.0010}                 & \erm{0.1813}{0.0017}{0.0013}                 & \erm{0.1812}{0.0018}{0.0011}                 & \erm{0.1814}{0.0018}{0.0013}                 \\
$r_{\rm b}$           & \mc{ }                                       & \erm{0.02791}{0.00053}{0.00011}              & \erm{0.02794}{0.00047}{0.00011}              & \erm{0.02794}{0.00047}{0.00014}              & \erm{0.02797}{0.00053}{0.00017}              \\
$\sigma$ ($m$mag)     & \mc{ }                                       & \mc{ 0.8518}                                 & \mc{ 0.8517}                                 & \mc{ 0.8517}                                 & \mc{ 0.8518}                                 \\
$\chi^2_{\rm \ red}$  & \mc{ }                                       & \mc{ 1.1364}                                 & \mc{ 1.1362}                                 & \mc{ 1.1364}                                 & \mc{ 1.1365}                                 \\
\hline \hline \end{tabular} \end{table*}

\end{document}